\documentstyle[aps,epsfig,cite]{revtex}
\newcommand{\mylims}{\stackrel{\scriptscriptstyle N\rightarrow\infty}
{\longrightarrow}}

\begin{document}

\title{\vspace*{-20mm} Clones and other interference effects in the 
evolution of angular momentum coherent states}
\author{ P. Rozmej \cite{EMroz} }
\address{Theoretical Physics Department, Institute of Physics,\\
University Maria Curie-Sk\l odowska, PL 20-031 Lublin, Poland\\
and \\ Institut des Sciences Nucl\'eaires, F 38026 Grenoble-Cedex,
France}
\author{R. Arvieu \cite{EMarv} }
\address{Institut des Sciences Nucl\'eaires, F 38026 Grenoble-Cedex,
France}
\date{Received 13 January 1998, in revised form 10 July 1998}
%
  \maketitle
 \begin{abstract}
  The aim of this article is to present the interference effects which
occur during the time evolution of simple angular wave packets (WP) which 
can be associated to a diatomic rigid molecule (heteronuclear) or to a 
quantum rigid body with axial symmetry like a molecule or a nucleus. The 
time evolution is understood entirely within the frame of fractional 
revivals discovered by Averbukh and Perelman since the energy spectrum is 
exactly quadratic. Our objectives are to study how these interference 
effects differ when there is a change of the initial WP. For this purpose 
we introduce a two parameter set of  angular momentum coherent 
states. From one hand this set emerge quite naturally from the three 
dimensional coherent states of the harmonic oscillator, from another hand 
this set is shown to be buit from intelligent spin states. By varying one 
parameter ($\eta$) a scenario of interferences occur on the sphere at 
fractional times of the revival time that strongly depends on $\eta$. 
For $\eta=\pm 1$ the WP,
which coincides with a WP found by Mostowski, is a superposition of Bloch 
or Radcliffe's states and clone exactly in time according to a scenario 
found for the infinite square well in one dimension and also for the two 
dimensional rotor. In the context of intelligent spin states it is 
natural to study also the evolution by changing $\eta$. 
For $\eta=0$ the WP is 
called linear and produces in time a set of rings with axial symmetry 
over the sphere. The WP for other values of $\eta$ are called elliptic and   
sets of fractional waves are generated which make a transition between 
two symm etries. We call 'mutants' these fractional waves. For specific 
times a clone is produced that stands among the mutants. Therefore the 
change in $\eta$ produces novel change in the quantum spread on the 
sphere. We have also constructed
 simple coherent states for a symmetric rotor
 which are applicable to molecules and nuclei.
 Their time evolution also shows a cloning
 mechanism for the rational ratio of moments of inertia. For irrational
 values of this ratio, the scenario of partial revivals completed
 by Bluhm, Kostelecky and Tudose is valid.

 \end{abstract}


\section{Introduction} \label{sec1}

 In the recent years one has found the existence of a generic behaviour 
for the time evolution of simple quantum systems.
In ref.\  \cite{averbukh} Averbukh and 
and Perelman have indeed discovered  a universal scenario of
fractional revivals in the long term evolution of quantum wave
packets of bounded system which goes beyond the correspondence
principle. They established this scenario by expanding the bound
states energies relevant to the wave packet up to the second order
with respect to the mean energy, thus producing a local
spectrum linear plus quadratic in one quantum number.
For a well concentrated wave packet at the initial time they defined
two time constants $T_{cl}$ and $T_{rev}>T_{cl}$ such that the
foregoing evolution of the wave packet is predicted as follows:
for $0<t<T_{cl}$, the wave packet spreads around the mean trajectory
that can be associated with the underlying classical evolution
while for $T_{cl}<t<T_{rev}$ the wave packet interferes with itself
in such a way that for fractional times $t=(m/n)\,T_{rev}$ the
wave packet is divided into $q$ fractional wave packets. If $n$ is
even, $q=n/2$, if not, then $q=n$. The specificity of the system
and that of the wave packet determines the shape of the fractional
wave packets
which are supposed to be spread regularly around the mean trajectory.
At time $t=T_{rev}$, the wave packet is rebuilt either identical
or only resemblant to the original one, depending on the importance
of the neglected terms higher than quadratic. In some very specific
cases, the fractional wave packets are clones of the initial one
as studied
in the most recent paper \cite{aronstein} and the revival is exact.
This scenario has been validated in the most spectacular manner by
several authors for a wave packet in a circular orbit of the
hydrogen atom \cite{dacic}, see also \cite{nauenberg} and by
Boris et al.
\cite{boris} for an elliptic orbit. It has been extended by Bluhm
and Kostelecky \cite{bluhm} to a very long evolution with the
demonstration of superrevivals due to the cubic terms.
The analytical explanation of the effects caused by cubic terms has 
been found by Leichtle et.al.\ \cite{leichtle}.  
Examples of vibrational wave packets in anharmonic potential of
simple molecule were also found since \cite{bowman,vrakking}.
The scenario has been thereafter extended to cases where the energy
depends on two quantum numbers \cite{bluhm1}.
A recent synthesis \cite{bluhm2} contains most of the 
references on this topic while ref.\ \cite{alber}
has been devoted to the definition of the experimental
of wave packets in atomic physics and in molecular physics.
 The recurrences studied in 
ref.~\cite{averbukh,aronstein,dacic,nauenberg,boris,bluhm,leichtle,
bowman,vrakking,bluhm1,bluhm2,alber} 
are all devoted to wave packets 
and can be mathematically explained in terms of special Gauss'sums.
It is necessary to point out that such sums were earlier extensively 
used in the physical litterature by Berry and Golberg \cite{berrygold},
and Berry \cite{berryphd} who studied the full time evolution of the 
propagator of a nuclear spin with an hamiltonian of the form $l_z^2/2J$.
These authors developed a renormalization theory and discussed 
the semiclassical limit.
More recently \cite{berryklein} various Talbot effects 
(integer, fractional, fractal) were also 
discovered with the help of these techniques.
Finally Berry \cite{berryjpa} has shown the occurrence of fractal 
dimensions both in space and time during the 
evolution of a uniform WP in boxes of arbitrary dimensions.

 Our aim is to consider only angular wave packets for diatomic molecules
and also for symmetric rotors and our purpose is to discuss the time
evolution of a large enough set of coherent states of the angular 
momentum. There is rather extensive literature on the coherent states. 
A full bibliography on this subject is found in ref. \cite{klauder}.
But we will focus on the {\em intelligent spin states} described in 
ref.\ \cite{bloch,aragone1,aragone2,rashid,kolod,vetri}
Large efforts have been made to build up such wave packets mainly in
the context of group theory. Little effort, on the contrary, has been
made to understand in detail their time evolution.
Until the work of \cite{averbukh} it was not realized that
they could evolve according to an universal scenario. For a diatomic
molecule or for a symmetric rotor, the spectrum of which are quadratic
in quantum number like for the systems discussed in
\cite{aronstein,vetchinkin},
the universal scenario is exact and repeats with
period $T_{rev}$. The main question is whether the fractional wave
packets are clones or just resemblant to the initial ones.
Despite all efforts to concentrate initially the wave packet in the
best possible way, the quantum evolution destroys this concentration
(spatial localization) according to the rule formulated in
\cite{averbukh}. We will show in this paper under which conditions
the wave packets separate into clones and in which conditions there is
a more restricted scenario of partial revivals.
We will also show the existence of new fractional waves with
different shapes that we call mutants.
It is crucial to present
a simple and physically meaningful picture of a coherent state and to
decrease the number of its parameters to the minimum. One should keep
in mind that a coherent wave packet has a classical content larger than
the eigenstates of the angular momentum. Due to this property, it is
possible to choose in the simplest manner the coherent wave packet
as we will show below.

 In section \ref{sec2} we will study a large set of coherent angular WP which 
depend only on the angles $\theta$ and $\phi$. If we impose one condition of 
minimum uncertainty these WP are composed of eigenstates of $L^2$ and 
contain, after a proper choice of the axis of coordinates, only spherical 
harmonics with magnetic quantum number $m$ of the same parity as $l$, 
These states belong to the family of coherent states called 
the {\em intelligent spin states} 
\cite{bloch,aragone1,aragone2,rashid,kolod,vetri}.
The main body of this article is organized around a particular subset: 
the exponential WP which are shown to be narrowly related to the coherent 
states of the harmonic oscillator (see the Appendix). Moreover the angular 
spread of the probability density depends on a single adjustable parameter.
One of the limiting case is a WP derived originally by Mostowski 
\cite{mostowski} for a diatomic molecule and is called circular. 
By varying one parameter one obtains new WP with cylindrical symmetry that 
we call linear and a large set called elliptic. 
Each element of this set corresponds to a quantum system 
like a diatomic molecule or a nucleus in a pure state with a specific 
preparation giving to it an average angular momentum, an angular 
distribution and particular spread of the distribution of the angular momentum. 

   In section \ref{sec3} it will be shown that a WP obtained by 
Atkins and Dobson \cite{atkins} using Schwinger \cite{schwinger}
boson representation of spin 1/2 is a
particular circular state almost coincident with the exponential WP.
More generally the boson representation for a spin $s$ is also shown to
lead to circular states. The exponential WP provides a closed and compact
expression in the angular variables. 

   In section \ref{sec4} we will state precisely for our angular WP the scenario 
of fractional revivals derived in \cite{averbukh} for the general case. 
It is indeed possible to specify the shape of the fractional revivals. 
The most 
spectacular event of cloning found essentially for the infinite square 
well in one dimension \cite{aronstein}  
and for a two dimensional rotor in \cite{bluhm2} is 
extended here to the most general case of circular WP. 
For the linear and 
elliptic WP the fractional waves are generally different from the initial WP
due to the quantum spread. However for a particular set of times a single 
clone exists which coincides with the initial WP. In the case of the most 
general linear WP the fractional waves keep cylindrical symmetry showing 
isotropy in the spread over the sphere. For the elliptic WP the fractional 
waves should accomodate two limiting symmetries: the plane symmetry 
present for circular states and the cylindrical symmetry valid for the 
linear states. We call 'mutants' these intermediate fractional waves.

   In section \ref{sec5} we present a numerical calculation showing 
this evolution for exponential WP.
The cases where cloning occurs is somewhat
obvious, however it is interesting to define properly the time windows
during which a given system of clones governs the time evolution. The
'carpet' representation used also elsewhere \cite{grossmann} 
is an interesting tool in this respect. 
A second interesting result of this section lies in the shape
of the mutants which can hardly be found from analytical considerations.
It is found that these fractional waves preserve a good angular localization
on the sphere. Their shape differs however from that of the initial WP
since the generic structure is a well defined crescent-like shape. 
    
Finally in section \ref{sec6}, we will construct an angular momentum
coherent state of a symmetric rigid rotor according to the rules defined
by Janssen \cite{janssen}. The time evolution of such a state
is studied in section \ref{sec7}.
Once the number of parameters is reduced,
the time evolution of the coherent state, which is now a three
dimensional system with two quantum numbers as in \cite{bluhm1},
presents clones if the ratio of the moments of inertia is rational.
In the case when this ratio is not a rational number the fractional wave
packets are not clones.

\section{Derivation of coherent angular wave packets} \label{sec2}
 
 Coherent angular WP can be defined as functions of $\theta$ and 
$\phi$ which  fulfill two requirements: 
\begin{enumerate}
\item Their angular spread should be under control i.e.\ it should be possible
to adjust their angular distribution in the easiest manner by changing a
few parameters or a simple function. 
\item A criterion of minimum uncertainty should be obeyed in a manner 
similar to the conditions satisfied by the coherent states of the 
harmonic oscillator.
\begin{equation} \label{c1}
   \Delta q_i \, \Delta p_i = \frac{\hbar}{2}, \hspace{5ex} i=x,y,z \;.
\end{equation}
 In that direction many attempts have used uncertainty relations derived
with the angular variables and the angular momentum operator ( see for
example \cite{klauder} for a complete reference to these works). Despite the long
list of works devoted to this field there is no detailed evolution of the
time evolution of these WP in the literature as said our introduction.
The first attempts in the line of the modern developments has been done in
\cite{bluhm2}
in which a WP of the rigid rotor in 2 dimensions was shown to clone
exactly. Our work will extend these results to three dimensions and will
discuss a rather large class of WP. 
\end{enumerate}
   In the following we will use only uncertainty relations based upon the 
components of the angular momentum and we will call coherent states all 
the states which satisfy 
\begin{equation} \label{c2}
      \Delta L_x^2 \, \Delta L_y^2 = \frac{1}{4} \langle L_z \rangle^2 \;.
\end{equation}
We assume that this equation holds with the axis of coordinates such that 
\begin{equation} \label{c3}
     \langle L_x \rangle=0 \, , \hspace{5ex}  \langle L_y \rangle=0 \;.   
\end{equation}
It is a textbook result that a general WP satisfies the inequality (in 
units with $\hbar=1$)
\begin{equation} \label{c4}
     \Delta L_x^2 \, \Delta L_y^2 \geq \frac{1}{4}|\langle [L_x,L_y] \rangle|^2 \;. 
\end{equation}
This result is derived by considering the norm of the state obtained by 
application of a special combination of $L_x$ with $L_y$ which involve a real 
parameter called $\eta$
\begin{equation} \label{c5}
     (L_x+ \mbox{i} \eta L_y)|\Psi\rangle  \;.  
\end{equation}
 If the minimum uncertainty condition is realized there exists a value of 
$\eta$ for which 
\begin{equation} \label{c6}
     (L_x+\mbox{i} \eta L_y)|\Psi\rangle=0 \; . 
\end{equation}
This value of $\eta$ is related to the average values by two formulas
\cite{jackiw} 
\begin{equation} \label{c7}  
   \eta= \frac{\langle L_z \rangle}{2\Delta L_y^2} =
   \pm \sqrt{\frac{\Delta L_x^2}{\Delta L_y^2}} \;. 
\end{equation}
The second of these equations provides a meaningful interpretation of $\eta$ 
in terms of $\Delta L_x^2$ and $\Delta L_y^2$. Let us now construct states which 
satisfy eq.\ (\ref{c6}).

\subsection{Eigenstates of $L^2$.} \label{ss2a}

  The simplest and most natural possibility is to construct the states 
$|\Psi\rangle $ as eigenstates of $L^2$.
This problem was solved long ago 
\cite{bloch,aragone1,aragone2,rashid,kolod,vetri} and the 
solutions were called {\em intelligent spin states}. Let us briefly sketch a few 
of their properties and explain why we will consider only a subset 
of them.
 
  The intelligent spin states are the eigenstates $|w\rangle $ of $L^2$ 
and of the (non hermitian) operator $L_x+\mbox{i} \eta\,L_y$ 
with eigenvalue $w$ such that
\begin{equation}\label{c8a}
      (L_x+\mbox{i}\eta\, L_y)\,|w\rangle =w\,|w\rangle  \; . 
\end{equation}
	The $(2l+1)$ eigenvectors of (\ref{c8a}) are discussed
extensively in \cite{aragone1,aragone2,rashid}. 
It has been shown by Rashid \cite{rashid} that there is a one
to one correspondance between each eigenvector $|w\rangle$  and a parent state
$|lm\rangle$. Therefore instead of $|w\rangle$ it is better to denote a 
solution of (\ref{c8a}) as $|\eta \,l m\rangle$. 
The relation between $|lm\rangle$  and $|\eta\, lm\rangle$ implies a 
normalisation factor $a_{lm}$ given in \cite{rashid} and an operator such that  
\begin{equation}\label{c9a}
 |w\rangle =|\eta\,lm\rangle = a_{lm}\, \exp(\delta\,L_z)
 \,\exp(-\mbox{i}\frac{\pi}{2} \,L_y)\, |lm\rangle  \; . 
\end{equation}
   The parameter $\delta$ is related to $\eta$ by 
\begin{equation}\label{c10a}
   \exp(\delta)=\sqrt{\frac{1+\eta}{1-\eta}}  \; , 
\end{equation}
 and the eigenvalue $w$ is expressed in terms of $m$ by
\begin{equation}\label{c11a}
               w = m\, \sqrt{1-\eta^2}   \; . 
\end{equation}
Among the $2l+1$ states (\ref{c9a}) one can identify: 
\begin{enumerate} 
  \item The states with $m=0$ which fulfill eq.\ (\ref{c2}) and (\ref{c3}).
  These states will be taken into account and will be explicited thereafter.
  \item The states with $m=\pm 1$. These particular states were first 
introduced by Bloch and Radcliffe \cite{bloch}.
In a more convenient system of 
coordinates they fulfill the simpler equation with $\eta=1$
\cite{aragone2,rashid}
\begin{equation}\label{c12a}
           (L_x+\mbox{i}\, L_y)\,|w\rangle =0   \; . 
\end{equation}
 Therefore these states will also be considered in our paper and we will 
call them circular states.             
  \item The states for which the parent value $m$ is neither 0 or $\pm1$. 
These states do not coincide with the previous ones. However they are
not orthogonal to them. Moreover they require a value of $l$ larger or equal
to 2 i.e.\ a tensor of rank at least equal to two is needed in order to
generate them. We have not studied these states and it is still an open
question to build a convenient WP by implying them. 
\end{enumerate} 
On the contrary the states with $m=0$ can be generated quite naturally starting 
from a three dimensional gaussian WP as shown in the Appendix and require 
a very simple vector operator. 
The states with $m=0$ have a very simple structure in terms
of spherical harmonics that is worth to be presented shorly independently
on the general solution found in \cite{rashid}.
Let us denote by ${\cal Y}^{l}_{\eta}(\theta,\phi)$ 
these new spherical harmonics which depend on a continuous real parameter 
$\eta$ and let us expand them in terms of the usual $Y^l_m$ as
\begin{equation} \label{c8}   
{\cal Y}^{l}_{\eta}(\theta,\phi)=\sum_{m=-l}^{l} C^l_m(\eta) Y^l_m(\theta,\phi) \;.
\end{equation}
The recurrence between the $C^l_m$ derived from eq.\ (\ref{c6}) 
implies that the sum (\ref{c8}) is restricted in such a way that 
$m$ and $l$ have the same parity, indeed the recurrence is: 
\begin{equation} \label{c9}
  C^l_{m+1} = -  C^l_{m-1}\, \frac{1+\eta}{1-\eta} \, 
  \sqrt{\frac{l(l+1)-m(m-1)}{l(l+1)-m(m+1)}} \;.
\end{equation}
In the following we will need the expression of 
${\cal Y}^l_\eta$ for $l=1$ which is
\begin{eqnarray} \label{c10}
 {\cal Y}^{1}_{\eta}(\theta,\phi) & = &
 \frac{(1+\eta)Y^1_1-(1-\eta)Y^1_{-1}}{\sqrt{2(1+\eta^2)}} \\
 & = & -\frac{1}{4}\,\sqrt{\frac{3}{\pi}}\,\frac{1}{\sqrt{1+\eta^2}} \,
 \sin \theta(\cos\phi +\mbox{i}\eta\sin\phi) \nonumber \;.
\end{eqnarray}
 The combination of $\theta$, $\phi$ and $\eta$ given above will be defined as
\begin{equation} \label{c11}  
           v=\sin\theta(\cos\phi+\mbox{i}\eta\sin\phi) \;. 
\end{equation}
One has 
\begin{equation} \label{c12}
 \langle {\cal Y}^{1}_{\eta} | L_z | {\cal Y}^{1}_{\eta} \rangle =
  \frac{2\eta}{1+\eta^2} \;. 
\end{equation}
Similarly, for $l=2$ the coherent states are
\begin{equation} \label{c13}  
{\cal Y}^{2}_{\eta}(\theta,\phi) = \left[ \frac{8}{3}(1+\eta^4)+
\frac{32}{3}\eta^2\right] ^{-1/2} \, 
\left\{ (1+\eta)^2 Y^2_2-\sqrt{\frac{2}{3}}(1-\eta^2)Y^2_0
+(1-\eta)^2 Y^2_{-2} \right\} \;, 
\end{equation}
while the average of $L_z$ is
\begin{equation} \label{c14}
 \langle {\cal Y}^{2}_{\eta} | L_z | {\cal Y}^{2}_{\eta} \rangle = 
 \frac{6\eta(1+\eta^2)}{1+4\eta^2+\eta^4} \;. 
\end{equation}
It is interesting to point out that $\eta=\pm 1$ corresponds to states with 
$m=\pm l$ while the states with $l=0$ are eigenstates of $L_x$ and can be the 
more simply written as single spherical harmonics of the angle $\theta'$ 
defined as
\begin{equation} \label{c15}  
          \cos\theta'=\sin\theta\cos\phi  \;,
\end{equation}
and one has   
\begin{equation} \label{c16}
 {\cal Y}^{l}_{0}(\theta,\phi) =  Y^{l}_{0}(\theta',\phi') = 
 \sum_{m=-l}^{l} C^l_m(0)\; Y^l_m(\theta,\phi) \;.  
\end{equation}
 
  The coherent spherical harmonics ${\cal Y}^{l}_{\eta}(\theta,\phi)$ 
have no freedom in them which allows a proper angular localization. 
It is therefore necessary to consider linear combinations. 

\subsection{General WP.} \label{ss2b}
 
  The most general WP solutions of (\ref{c6}) which fulfill eq.\
  (\ref{c2}) will be written as
\begin{equation} \label{c17} 
 |\Psi_{\eta}\rangle_{\mbox{\scriptsize general}} = \sum_l \, \lambda^{l}_{\eta} \, 
 | {\cal Y}^{l}_{\eta}\rangle \;.  
\end{equation}             
They depend on $\eta$ which can be interpreted with the help of eq.\
(\ref{c7}) and on weights $\lambda^{l}$ 
which can be determined in order to provide a convenient 
angular localization. Again there will be states with $\eta=\pm 1$ that 
will be called circular and others with $\eta=0$ that will be called 
linear. The WP defined with other values of $\eta$ will be called elliptic in 
the following. The justification of this name will be given in the Appendix.
    In section \ref{sec3} many results will be derived for the 
particular class of WP that will be defined in the following subsection. 
Most of them can be seen to apply to the general WP (\ref{c17}).

\subsection{Exponential coherent WP.} \label{ss2c}

   Among functions that are possible those which can be 
expanded in a power series of the variable $v$ defined by (\ref{c11}) are 
particularly interesting because they will contain all the partial waves 
${\cal Y}^l_\eta$. We have chosen to concentrate on the exponential coherent WP
\begin{equation} \label{c18}
  \Psi_{\eta} (\theta,\phi) = \sqrt{\frac{N}{2\pi\sinh 2N}} \,
  \mbox{e}^{N\sin\theta(\cos\phi+\mbox{\scriptsize i}\eta\sin\phi)} \;, 
\end{equation}
which possess important properties: they fulfil eq.\ (\ref{c2}), they
have direct connection to coherent states of harmonic oscillator
(see the Appendix) and finally they have a simple geometrical interpretation.
Indeed the real parameter $N$ introduced there allows a proper adjustment 
of the angular spread. 
The probability density depends only on $N$ and on the angle 
$\theta'$ defined by (\ref{c15}), the expression is
\begin{equation} \label{c19}
 |\Psi_{\eta} (\theta,\phi)|^2 = \frac{N}{2\pi\sinh 2N}
 \,\mbox{e}^{2N\cos\theta'}  \;. 
\end{equation}
 If we put $\eta=1$ into (18) we obtain a coherent state defined by 
Mostowski \cite{mostowski} who wrote it as 
\begin{equation} \label{c20}  
 \Psi_M(\theta,\phi)= C^{-\frac{1}{2}} \, 
 \mbox{e}^{N(\vec{u}_1+\mbox{\scriptsize i}\vec{u}_2) \cdot \vec{n}} \;.  
\end{equation}
Here $\vec{u}_1$ and $\vec{u}_2$ are two perpendicular unit vectors 
(in our case $\vec{u}_1$ is along $Ox$ and $\vec{u}_2$ along $Oy$ 
and we have eq.\ (\ref{c3}) and $\vec{n}$ is a unit vector in the 
direction $(\theta,\phi)$.

  The generalization of (\ref{c20}) with a parameter $\eta$ was never considered 
until now to our knowledge and the time evolution was never studied yet.
  Let us point out that eq.\ (\ref{c18}) can be generalized as 
\begin{equation} \label{c21}
 \Psi_{\eta} (\theta,\phi) = C^{-\frac{1}{2}}\, 
 \mbox{e}^{N(\vec{u}_1+\mbox{\scriptsize i}\eta\vec{u}_2)\cdot \vec{n}}    
\end{equation}
 with arbitrary but perpendicular $\vec{u}_1$ and $\vec{u}_2$.
 Calling $\vec{u}_3$ a third unit vector perpendicular to 
$\vec{u}_1$ and $\vec{u}_2$ we will obtain WP which do not 
fulfill eq.\ (\ref{c2}) but rather 
\begin{equation} \label{c22}  
 \Delta L_1^2 \, \Delta L_2^2 = \frac{1}{4}\langle L_3 \rangle^2 \;. 
\end{equation}
Instead of eq.\ (\ref{c3}) we would have 
\begin{equation} \label{c23}  
       \langle L_1 \rangle=0 \;, \hspace{5ex}   \langle L_2 \rangle=0 \;. 
\end{equation}
  The choice of axis made in eq.\ (\ref{c18}) simplifies much the 
interpretation and the partial wave expansion. 
This expansion will be given in subsection \ref{ss2d}.

  It is not difficult to derive that our $\Psi_\eta$ defined by (\ref{c18}) 
have the following average values and the following limits for large $N$:
\begin{equation} \label{c24}
 \langle\Psi_\eta | L_z | \Psi_\eta\rangle =  \langle L_z \rangle =
 \eta[N\coth(2N)-\frac{1}{2}] \,\mylims \,
 \eta[N-\frac{1}{2}]  \;. 
\end{equation}
Therefore eq.\ (\ref{c7}) implies that 
\begin{equation} \label{c25}
 \Delta L_y =  \langle L_y^2 \rangle = \frac{1}{2}[N\coth(2N)-\frac{1}{2}] \,
 \mylims \, \frac{1}{2}[N-\frac{1}{2}] \;,
\end{equation}
\begin{equation} \label{c26}
 \Delta L_x =  \langle L_x^2 \rangle = \frac{\eta^2}{2} [N\coth(2N)-\frac{1}{2}] \,
 \mylims \, \frac{\eta^2}{2}[N-\frac{1}{2}] \;.
\end{equation}
The average of $L_z^2$ and the total uncertainty $\Delta L_{\eta}^2$
need independent calculations with the results:
\begin{equation} \label{c27}
 \langle L_z^2 \rangle = \langle L_y^2 \rangle [1-2\eta^2] + \eta^2 N^2  \;, 
\end{equation}
\begin{equation} \label{c28}
 \Delta L_{\eta}^2 = \langle L^2 \rangle - \langle L_z^2 \rangle 
 \, \mylims \, N- \frac{1}{2} +\eta^2 \,\frac{N}{2}  \;. 
\end{equation}

\subsection{Partial wave expansion of the exponential coherent state.}\label{ss2d}
  
  We need to distinguish the case with a general value of $\eta$ from the 
simpler cases with $\eta=1$ or 0. (The cases with $\eta =-1$ or negative $\eta$
 are trivially deduced fom the cases with positive $\eta$ by inversing the sense 
of rotation of the WP).
 For $\eta=1$, i.e.\ circular exponential WP, or Mostowski's WP one has
\begin{equation} \label{c29}  
 \Psi_M(\theta,\phi) = \Psi_1(\theta,\phi) = \sqrt{\frac{2N}{\sinh 2N}} \,
 \sum_{I=0}^{\infty} \, \frac{(2N)^I}{\sqrt{(2I+1)!}} \, Y^I_I(\theta,\phi) \;.
\end{equation}
 For $\eta=0$, i.e.\ linear exponential WP, it is found that
\begin{eqnarray} \label{c30} 
 \Psi_0(\theta,\phi) & = & \Psi_0(\theta',\phi') = \sqrt{\frac{N}{2\pi\sinh 2N}} \,
 \mbox{e}^{N\cos\theta'} \\  
& = & \sqrt{\frac{2N}{\sinh 2N}} \,\sum_{I=0}^{\infty} \, \sqrt{2I+1} \,
  \sqrt{\frac{\pi}{2N}} \, I_{I+\frac{1}{2}}(N) \, Y^I_0(\theta',\phi')
 \label{c31}  \;,
\end{eqnarray}
 where $I_{I+\frac{1}{2}}(N)$ is a spherical Bessel function of the first kind.
 
 For a general value of $\eta$, i.e.\ elliptic WP, the argument $v$ can be 
separated into two parts
\begin{equation} \label{c32}
 v = \frac{1+\eta}{2}\,\sin\theta \,\mbox{e}^{\mbox{\scriptsize i}\phi} 
   + \frac{1-\eta}{2}\,\sin\theta \,\mbox{e}^{\mbox{\scriptsize -i}\phi} \;. 
\end{equation}
Then $\mbox{e}^{Nv}$  
is calculated as two power series containing products of $Y^l_l$ 
and $Y^{l'}_{-l'}$. These products are then expanded in terms of $Y^I_M$  as
\begin{equation} \label{c33}  
\Psi_{\eta}(\theta,\phi) =  \sum_{IM}\, b_{IM}(N,\eta) \,
Y_M^I(\theta,\phi)  \, ,  
\end{equation}
 with the weights $b_{IM}$ given by 
\begin{equation} \label{c34}  
b_{IM}(N,\eta) = \sqrt{\frac{2N}{\sinh (2N)}} \, \sum_{l l'} \,
\frac{(-1)^{l'} [N(1+\eta)]^l \,[N(1-\eta)]^{l'}}{\sqrt{(2l)!(2l')!}} \,
\frac{\langle ll'00|I0\rangle
\langle ll'l-l'|IM\rangle}{\sqrt{2I+1}} \; . 
\end{equation}

     From the discussion made in subsection \ref{ss2a} $b_{IM}$ 
is proportional to $C_{IM}$.
The known selection rules of the Clebsh-Gordan coefficients which 
appear in (\ref{c34}) assure that $M$ should have the parity of $I$.

\section{Comparison with coherent states defined in terms
of bosons} \label{sec3}

  In this section we will compare the previous coherent states to an 
other set defined in terms of bosons of spin $s$.
Based on Schwinger's work \cite{schwinger} several angular momentum
coherent states have been constructed which rely upon boson
representation of spin $s$. The case with $s=1/2$ was first
considered by Bonifacio et al. \cite{bonifacio} and studied
more extensively by Atkins and Dobson \cite{atkins}.
Mikhailov \cite{mikhailov} has generalized this case to any spin
integer or half integer. His work was complemented by Gulshani
\cite{gulshani}. According to Mikhailov, a coherent state formed
with $2s+1$ bosons depends on 2 complex numbers called
$\alpha_+$ and $\alpha_-$ combined to define other complex
constants $\alpha_{jm}$ by
\begin{equation}\label{c35}	    
 \alpha_{jm} = {\alpha_+}^{j+m} {\alpha_-}^{j-m}
 \left ( \begin{array}{c} 2j\\ j-m \end{array}
 \right )^\frac{1}{2}  \, ,
\end{equation}
for $j=0,s,2s,\ldots,ps,\ldots$ and $m=-j,-j+1,\ldots,j-1,j$.

The coherent state, called generically $|\alpha s\rangle$,
is expressed by
\begin{eqnarray}\label{c36}    
|\alpha s\rangle & = & \exp(-\frac{n^{2s}}{2}) \,\prod_{\mu} \,
  \exp(\alpha_{s\mu} a_{\mu}^\dagger) \, |0\rangle \\
  \label{c37} 		   
 & = & \exp(-\frac{n^{2s}}{2}) \, \sum_{j=0,s,\ldots,ps,\ldots}^{\infty}\,
  \sum_{m=-j}^{j} \,\frac{1}{\sqrt{p!}} \,\alpha_{jm}\, |jms\rangle \, ,
\end{eqnarray}
$|0\rangle$ represents the vacuum, $a_{\mu}^\dagger$
($\mu=-s,\ldots,s$) is a creation operator of a boson of spin $s$,
while the normalization constant depends on $\alpha_+$ and $\alpha_-$
through
\begin{equation}\label{c38}	 
 n^{2s} = (|\alpha_+|^2 + |\alpha_-|^2)^{2s} = \sum_{\mu}
|\alpha_{s\mu}|^2  \, .
\end{equation}
The states $|jms\rangle$ are normalized states of angular momentum
$j$ constructed from the $a_{\mu}^\dagger$ [see Mikhailov for the
full expressions]. The expansion (\ref{c37}) contains integer and
half integer $j$ for $s=1/2$, it reduces to integer $j$ for $s=1$,
for $s=2$ only even $j$ occurs etc.
The states (\ref{c36})--(\ref{c37}) are 
eigenstates of the annihilation operator
$a_{\mu}^\dagger \hspace{2ex} (\mu=-s\ldots +s)$ with eigenvalue 
$\alpha_{s\mu}$.
Mikhailov has also calculated the expectation values of various
operators which are expressed in terms of $a_{\mu}^{\dagger}$ and
$a_{\mu}$. Due to this technical simplification, the calculations of
expectation values are easier than the work necessary to obtain
formulas (\ref{c24}) to (\ref{c28})
in the case of the coherent state $\Psi_{\eta}(\theta,\phi)$.
It is interesting to
compare the properties of the state $|\alpha s\rangle$ to those of 
the general state defined by eq.\ (\ref{c17}) or to (\ref{c18}).
Since $\alpha_{jm}$ is generally non zero for all
values of $j$ and $m$ the states $|\alpha s\rangle$ 
cannot be identified with our
elliptic state for which $m$ should be of the same parity as $j$. 
Clearly formulas (\ref{c35}) and (\ref{c34}) are different.
At first sight one could think that (\ref{c36}) 
which depend on $s$ and on two complex numbers $\alpha_+$ and $\alpha_-$
describe an ensemble of WP larger than our WP (\ref{c18}) 
which depend only on $N$ and $\eta$ 
and would therefore apply to a larger variety of physical
situations. We will show that on the contrary the states (\ref{c36}) are a
particular set of circular states and do not contain the freedom
allowed with the parameter $\eta$. Indeed if $\alpha_+$ and $\alpha_-$ 
are both nonzero the states (\ref{c36}) 
do not fulfill the conditions (\ref{c3}). In order to
fulfill these conditions it is necessary that either 
$\alpha_+$ or $\alpha_-$ should be zero.
[In \cite{mikhailov} it is shown indeed that
    $\langle L_x \rangle \sim \hbox{Re}(\alpha_+^* \, \alpha_-), \hspace{2ex} 
     \langle L_y \rangle \sim \hbox{Im}(\alpha_+^* \, \alpha_-)$].
A change of axis leads to 
$\alpha_{jm}=0$ except if $m=j\/$ (if $\alpha_-=0$) or if $m=-j\/$ 
(if $\alpha_+=0$). 
The phase of $\alpha_+$ or $\alpha_-$ can be incorporated in the phase of the 
states $|j, m=\pm j, s\rangle $. Choosing for example $\alpha_-=0$ 
and $\alpha_+=k$ = real 
number the state $|\alpha s\rangle$ is now simply denoted as $|ks\rangle$
\begin{eqnarray}\label{c39}	  
|k s\rangle & = & \exp{[-\frac{(k^2)^{2s}}{2} ]}\,
\sum_{j=0,s,\ldots,ps,\ldots}^{\infty} \, \frac{1}{\sqrt{p!}} \,
(k^2)^j \, |j=ps,m=j,s\rangle  \\
\label{c40}			  
 & = & \exp(-\frac{k^{4s}}{2}) \, \exp[\alpha_{ss}
 a_{s}^{\dagger}] \, |0\rangle \, .
\end{eqnarray}

In the previous section, it was clear from the compact expression
(\ref{c20}) given by Mostowski that the same physical state
can be written by introducing three additional angles which
are necessary to specify $\vec{u}_1$ and $\vec{u}_2$.
The proof that this freedom exists also in the boson representation
was given by Mikhailov \cite{mikhailov}. By a convenient choice of
axis the state written in eq. (\ref{c37}) with $2s+1$ bosons
and two complex parameters can be brought to the simple form
(\ref{c39}-\ref{c40}) with only one boson of spin $s$ with $\mu=s$ and
one parameter $\alpha_{ss}= \alpha_+^{2s}=k^{2s}$.
Such a choice was also done by Atkins and Dobson \cite{atkins} for
the case $s=1/2$.
The matrix elements of the components of $\vec{L}$ given below
are a particular case of formulas given in ref. \cite{atkins}.
\begin{equation}\label{c41}	   
\langle ks | L_z | ks \rangle = s\, k^{4s} \, ,
\end{equation}
\begin{equation}\label{c42}	   
\langle ks | L_z^2 | ks \rangle = s^2\, k^{4s}(k^{4s}+1)  \, ,
\end{equation}
\begin{equation}\label{c43}	   
\langle ks | L_x^2 | ks \rangle = \langle ks | L_y^2 | ks \rangle \\
 = \frac{s}{2}\, k^{4s}   \, ,
\end{equation}
\begin{equation}\label{c44}	   
 \Delta L_x^2 \, \Delta L_y^2 = \frac{1}{4} \langle L_z\rangle^2 \, .
\end{equation}
These formulas and expansion (\ref{c39}-\ref{c40}) compared to formulas
(\ref{c24}) to (\ref{c27}) show that $|ks\rangle$ do not
coincide exactly with Mostowski's coherent state. For $s=1/2$
Atkins and Dobson have proposed to truncate the sum over $j$
in order to take into account only integer values of $j$. In this
process the new state that we will call $|k,\frac{1}{2}\rangle_i$
($i$ for integer) needs to be normalized properly and the formulas
(\ref{c41}) to (\ref{c43}) cease to be valid. Eq.\ (\ref{c44}) 
still holds because the state is nevertheless a circular state.

Our purpose is now to compare  $\Psi_M(\theta,\phi)$ to
$|k,\frac{1}{2}\rangle_i$. It is interesting to choose the
parameters $N$ and $k$ in such a way that the limit
(\ref{c24}) for large $N$ is valid ($\eta=1$). 
Then using (\ref{c41}) one puts
\begin{equation}\label{c45}	
\frac{1}{2}\,k^2 = N - \frac{1}{2} \, .
\end{equation}
The comparison presented in Fig.~1 
shows that the probability
${C_I}^2$ to find the partial wave $Y_I^I$ in $\Psi_M$ and in
$|k,\frac{1}{2}\rangle_i$ is practically the same for $N=20$.
For smaller $N$ and $k$ (i.e. $N$ of the order of unity)
we have observed small but not meaningful differences.
It is therefore possible, if $N$ is large enough, to identify
Mostowski's coherent state to the boson representation of spin 1/2
of Atkins and Dobson.

   We have shown in this section that the coherent states derived in 
the literature using a boson representation belongs to the particular 
class of coherent circular states defined in section \ref{sec2}.
The exponential coherent states defined in subsection
\ref{ss2c} above have the advantage of a well controllable 
localization and also depends on an interesting new parameter $\eta$.

\section{Time evolution of coherent angular WP} \label{sec4}

  The coherent states built in section \ref{sec2} are applicable 
to systems which have only angular coordinates on the sphere. 
This is the case of the three 
dimensional rotor with an axis of symmetry like an heteronuclear molecule 
or some deformed nuclei. In the following we will assume that the 
eigenvalues of these systems obey rigorously the $I(I+1)$ law and we will 
use the frequecy $\omega_0$ written in terms of the moment of inertia 
$J_0$ as
\begin{equation}\label{c46}	  
\omega_0 = \frac{\hbar}{2J_0}\, .
\end{equation}

    Our WP do not allow to consider other degrees of freedom 
like vibrational ones or internal excitations. However we are confident of 
the interest of our work which provides a full quantum-mechanical 
description of the rotation of a pure state of a three dimensional system.

    Let us define the {\em revival time}, $T_{rev}$ as 
\begin{equation}\label{c47}   
T_{rev}= \frac{2\pi}{\omega_0}=(2\bar{I}+1)\,T_{cl} \, .
\end{equation}
This time is twice the period of true revival of the WP. Indeed one has 
for the general case (\ref{c17})
\begin{equation}\label{c48}   
\Psi_{\eta}(\theta,\phi,t)_{general} =
\sum_I\; \lambda_I \,\mbox{e}^{-i\,2\pi I(I+1)\omega_0\,t}
{\cal Y}^I_\eta(\theta,\phi)                           
\end{equation}
Since $I(I+1)$ is always even the period is indeed $T_{rev}/2$.
Nevertheless we
will continue to use the same notations as in \cite{averbukh}.
These authors have also
introduced a second characteristic time $T_{cl}$ called the 
{\em classical time}. It
is defined in terms of the average angular momentum $\bar{I}$ 
defined by the average energy of the WP
\begin{equation}\label{c49}   
     \bar{I}(\bar{I}+1)= \sum_I \; |\lambda_I|^2 \, I(I+1)  
\end{equation}
For the case of an exponential WP one obtains with the help of 
eqs.\ (\ref{c25}-\ref{c27}):
\begin{equation}\label{c50}   
  \bar{I}(\bar{I}+1) = \langle L^2 \rangle
  \, \mylims \, \left ( N-\frac{1}{2}\right )
  +\eta^2 \left [ N^2-\frac{1}{2}(N-\frac{1}{2}) \right ]
\end{equation}
 The classical time $T_{cl}$ is defined as: 
\begin{equation}\label{c51}   
T_{cl}=\frac{2\pi}{\omega_0\, (2\bar{I}+1)} \, .
\end{equation}
$T_{cl}$ is the period of a classical rotator 
having angular momentum $I=\bar{I}$.
At times $t=(m/n)\,T_{rev}$ where $2m<n$ ($m$ and $n$ are mutually
prime integers) we will use the trick developed in  \cite{averbukh}
to write the quadratic exponential in $I$ as   
\begin{equation}\label{c52}   
 \mbox{e}^{-i\,2\pi\,I^2 \frac{m}{n}}   = \sum_{s=0}^{l-1} \; a_s \,
 \mbox{e}^{-i\,2\pi\, I \frac{s}{l}}               
\end{equation}
It is necessary to distinguish three cases:\begin{itemize}
\item[a)]
 $n$ is odd, then $l=n$ and all the coefficients $a_s$ are nonzero.
 However they have the same modulus $1/\sqrt{l}$. 
\item[b)]
 $n$ is even and multiple of four, then $l=n/2$ and the modulus of the $a_s$ 
 have the same value as above.
\item[c)]
 $n$ is even and not multiple of four, then $l=n$ but the $a_s$ with 
 even $s$ are zero, the other have their modulus equal to $1/\sqrt{n/2}$. 
\end{itemize} 
 The number of values of the $a_s$ which are nonzero is called $q$ 
 with $q=n$ if $n$ is odd and $q=n/2$ if $n$ is even. The phase of the 
 $a_s$ can be calculated as said in \cite{averbukh}. 
Using these results and inserting (\ref{c50}) into (\ref{c48}) one obtains
eqs.\ (\ref{c53}),(\ref{c54}) below:
\begin{eqnarray}\label{c53}   
\Psi_{\eta}(\theta,\phi,\frac{m}{n}T_{rev})_{general} & = &
\sum_I\; \lambda_I \, \sum_{s=0}^{l-1} \; a_s 
\mbox{e}^{-i\,2\pi (\frac{m}{n}+\frac{s}{l})}\; {\cal Y}^I_\eta(\theta,\phi)
\\ & = & \sum_{s=0}^{l-1} \; a_s 
 \Psi^s_{cl}(\theta,\phi,t_s) \; .  
 \label{c54}                        
\end{eqnarray}

  At times $t=(m/n)\,T_{rev}$
any WP is a sum of $q$ fractional WP $\Psi_{cl}^s$ each with a 
different effective time $t_s$
\begin{equation}\label{c55}   
         t_s=(\frac{m}{n}+\frac{s}{l})\,T_{rev}   
\end{equation}
  The fractional WP at times $t_s$ is given by
\begin{equation}\label{c56}   
 \Psi^s_{cl}(\theta,\phi,t_s) = \sum_I\; \lambda_I \,
 \mbox{e}^{-i\,I\omega_0\,t_s} \; {\cal Y}^I_\eta(\theta,\phi) \; . 
\end{equation}
There are several cases for which all the $\Psi_{cl}^s$ are clones of the 
initial WP defined by (\ref{c17}) for all possible valus of $t_s$. 
There are also cases where only one of all the fractional wave 
is a clone for particular $t_s$.
Let us describe now these events keeping in mind as far as possible 
arbitrary $\lambda_I$.

\subsection{Cloning of circular WP} \label{ss4a}

 If $\eta=1$ one has  
\begin{equation}\label{c57}   
\mbox{e}^{-i\,I\omega_0\,t_s} \; {\cal Y}^I_1(\theta,\phi) =
\mbox{e}^{-i\,I\omega_0\,t_s} \;  Y^I_I (\theta,\phi) = 
Y^I_I (\theta,\phi-\omega_0\,t_s)  \; .                         
\end{equation}
Independently of the $\lambda_I$ the fractional waves verify 
the cloning property
\begin{equation}\label{c58}   
 \Psi^s_{cl}(\theta,\phi,t_s) = \Psi(\theta,\phi-\omega_0\,t_s,0) \; .                          
\end{equation}
Among all circular WP which all clone in this way, the exponential WP, 
which can be sharply localized in the angle $\theta$ by considering high 
enough $N$, clone accordingly around $q$ directions disposed symmetrically in 
the Oxy plane defined by $q$ values of the angle $\omega_0\,t_s$. 

\subsection{Cloning for some particular $t_s$} \label{ss4b}
  
   This special situation occurs when there exist values of $s$ 
 such that $\omega_0\,t_s$ 
is a multiple of $2\pi$ and for which $a_s$ is non zero.
Let $s_0$ be defined by $s_0=n-m$.
  One has the property 
\begin{equation}\label{c59}   
 \Psi^{s_0}_{cl}(\theta,\phi,t_{s_0}) =  \Psi(\theta,\phi,0) \; .
\end{equation}
This event occurs whenever $n$ is odd or even and not multiple of four.
The clone is found always identical to the initial WP. Obviously it is 
multiplied by $a_{s_0}$. The existence of the clone is independent of the 
$\lambda_I$ and of $\eta$. 
  For the values of $s \neq s_0$ the fractional WP are different from the 
initial WP. Starting from $\eta=1$ their shape evolve with $\eta$ and
we propose to call them {\em mutants}. 
This {\em mutation} can indeed be seen as a transition 
between two symmetries as we will show numerically in section \ref{sec5}.

\subsection{Symmetry properties of the fractional waves} \label{ss4c}
   
   In the following discussion it will be assumed that $\lambda_I \neq 0$ 
both for even and odd values of $I$. 
If some further symmetry is assumed (for exemple $\lambda_I=0$ for odd $I$)    
new properties will result which will not be 
discussed in the present paper.

   The first remark is that, apart for the value $s_0$ defined above 
associated to a clone, the fractional waves can be paired for each value of 
$m$ and $n$ in the following manner: associated to $s$ there exists an other 
value $s'$ such that
\begin{equation}\label{c60}   
 \mbox{e}^{+i\,I\omega_0\,t_{s'}} =  \mbox{e}^{-i\,I\omega_0\,t_{s}} \; ,      
\end{equation}
 and also (noting a difference in the sign of $\eta$ in the right side)
\begin{equation}\label{c61}   
 \Psi^{s'}_{cl}(\theta,\phi,t_{s'})_{\eta} =  
 \Psi^{s}_{cl}(\theta,\phi,t_{s})_{-\eta}^{*} \; .                   
\end{equation}
This equation shows that the fractional waves coresponding to opposite 
$\eta$, i.e.\ opposite $\langle L_z \rangle$, are intermixed. 
   The equality (\ref{c61}) is based on the equality which defines $s'$ in terms 
of $s$
\begin{equation}\label{c62}   
\frac{t_{s}+t_{s'}}{T_{rev}} = 0 \hspace{1ex} \pmod{1}
\end{equation}
from one hand and from the conjugation property
\begin{equation}\label{c63}   
 {\cal Y}^I_{\eta}(\theta,\phi) =  {{\cal Y}^I_{-\eta}}^*(\theta,\phi)   \; .                    
\end{equation}
In addition to being real it is important that $\lambda_I$ should be an 
even function of $\eta$ (like for the exponential WP defined by 
(\ref{c33}-\ref{c34})).

  For $\eta=0$ the fractional waves $\Psi_{cl}^s$ and $\Psi_{cl}^{s'}$ 
 have the same probability density on the sphere.
This leads to a reduction in the number 
of fractional waves which occur: for odd $n$ there will be one clone plus 
$(n-1)/2$ fractional waves, for even $n$ and not multiple of four there is one 
clone and $(n-2)/4$ fractional waves, finally for $n$ multiple of four there 
will be $n/4$ fractional waves.
 
\section{Numerical calculations with exponential WP} \label{sec5}

    In this section we will describe some figures showing the time
evolution of typical exponential coherent WP. The value of $N$ will generally
be the same and we will change the parameter $\eta$.
Most of the figures are calculated for $N=20$.
This value is typical for a rather concentrated WP. 
Values near unity correspond to broad WP which occupy the whole of the
sphere and are not interesting for our purpose. Let us remind that keeping
the same $N$ and changing $\eta$ produces the same probability density 
(\ref{c19}) at $t=0$ however formula (\ref{c33}-\ref{c34}) shows that the 
distribution of the partial wave depends strongly on $\eta$.
The average $\bar{I}$ is very low for $\eta=0$ and this
produce a difference in the time evolution which shows less structure if
$N=20$ and if $\eta$ is decreased. These features can be seen if one study the
autocorrelation function represented for three values of $\eta$ in 
figure \ref{etaucor}. 
It is seen that this function is composed of peaks which have a
larger width if $\eta$ is small. The structure becomes very rich if $\eta$ is
nearer to 1. This autocorrelation function is very much similar to that 
studied in ref. \cite{vetchinkin}.  

 
 \subsection{Cloning for circular WP} \label{ss5a}
   
  The time evolution of the circular wave packet corresponding to $N=20$
is shown in Fig.~\ref{tevolM} and Fig.~\ref{carpetM}.
In Fig.~\ref{tevolM} a convenient set of times
have been chosen to show the probability density as a function
of  $\theta$ and $\phi$ just during the regime of spreading of
the wave packet ($t<T_{cl}$) then for a few cloning times followed
by the full revival for $t=T_{rev}/2$.
In Fig.~\ref{carpetM} a "carpet" is shown of the section
$0 \leq t \leq T_{rev}/2$ of the probability density for
$\theta=\pi/2$ which shows up to $q=7$ clones.

The difference between Mostowski's coherent state with that created
by Atkins and Dobson $|k,s=\frac{1}{2}\rangle_i$ is so tiny
that it does not present any interest to be shown.  
 Despite the analogy with the one dimensional results of the infinite 
square well \cite{aronstein} and the two dimensional rotor \cite{bluhm2} 
some new interesting 
aspects of our results need to be stressed. Indeed the circular WP 
spread in a way in the $\phi$ direction and clone around the $Oxy$ plane,
which is natural since there is a linear momentum along $Oy$ initially.
However there is no change in time in the azimuthal spread. 

The cloning mechanism 
found in quantum mechanics is not possible for a single classical particle 
however it will appear if one uses the ensemble interpretation of quantum 
mechanics as underlined by authors of ref.\ \cite{nauenberg}.

 \subsection{Linear WP }  \label{ss5b}                  
                                             
  For $\eta=0$ the fractional WP has cylindrical symmetry around $Ox$ at all
times since it is written as
\begin{equation}\label{c64}	
\Psi_{cl}(\theta',\phi',t) = \sum_{I} \, b_I \, e^{-2i\pi I\,t/T_{rev}}
\, Y^I_0(\theta',\phi') \, .
\end{equation}

The quantity $2\pi\sin \theta' |\Psi_{cl}|^2$ is represented in
Fig.~\ref{classL}. This shows that the revival wave packets are
rings on the  sphere and, due
to this special topology, do not clone the initial
wave function. 

The time evolution of $\Psi_{0}(\theta',\phi',t)$ is represented in
Fig.~\ref{carpetL}. Since the wave packet is in fact one dimensional, the
"carpet" representation provides the essential features of the
time evolution.
However 
  in order to produce a similar richness than for $\eta=1$ 
  (and a similar expectation value of $L^2$)
  we have increased the value of $N$ for $\eta=0$ to $N=50$.
On a sphere areas 
of constant probability density could be represented
by parallel circles centered on $Ox $ axis.
The pattern of the carpet shown in Fig.~\ref{carpetL}
is very much resemblant to that discussed recently for a special wave
packet in one dimensional box in ref.\ \cite{grossmann}.
Indeed there is a superposition of ridges and valleys with simple
slopes as a function of $t$. The interpretation of this effect can
be given in similar terms as in \cite{grossmann}. Note that
the quantity plotted in Figs.\ref{carpetL} and \ref{classL}
has the same boundary conditions as the wave packet on the edge of the box. 
This produces a reflection
effect in Fig.~\ref{carpetL} totally absent in Fig.~\ref{carpetM}
since the boundary conditions are different on the circle.
  Thus for $\eta=0$ the WP spreads uniformly in all directions defined by 
the angle $\phi'$. 

Such a WP is full nonsense for a single classical particle 
and takes sense only with the ensemble interpretation.

 Another interesting linear WP corresponding to 
$\eta\rightarrow\infty$ but keeping 
$\eta N$ finite is defined as follows: 
\begin{equation} \label{psietan}
        \Psi_{\eta N}(\theta,\phi) = \frac{1}{\sqrt{4\pi}} \,
	  \mbox{e}^{i\, \eta N \,\sin \theta\, \sin \phi}  \,. 
\end{equation}
 It is derived from the harmonic oscillator WP of the appendix by keeping 
only the term in $p_y$. This WP has its probability density uniformly 
distributed over the sphere and obeys the equation 
\begin{equation} \label{lypsi}
           L_y \, \Psi_{\eta N}=0    \,.   
\end{equation}
It has therefore cylindrical symmetry around $Oy$ and depends on the angle 
$\theta"$ defined by
\begin{equation} \label{teta"}
           \cos \theta"= \sin \theta \,\sin \phi  \,.   
\end{equation}
Its expansion in spherical harmonics contains now spherical Bessel functions
\begin{equation} \label{besetan}
 \Psi_{\eta N} = \sum_{I}\, \sqrt{2I+1}\,j_I(\eta N)\,Y^I_0(\theta",\phi") \,.  
\end{equation}
  Obviously the fractional waves have also cylindrical symmetry around $Oy$
but there is in addition, for $t=t_{s_0}$, 
a uniform clone which interferes with 
all the other fractional waves. The time evolution is shown in 
Fig.~\ref{evetan} for $\eta N=20$.
For the small values of $t$, like $t=1/100\,T_{rev}$, the WP is 
concentrated almost totally on the hemisphere
with $y>0$ with a spike along $Oy$ surrounded by concentric rings. 
For $t=1/50\,T_{rev}$ the same behaviour occurs 
but this time on the hemisphere with $y<0$. For other times both parts of the 
sphere are covered with rings and the spike also occurs on both sides of 
the $Oy$ axis. 
For $t=T_{rev}\,m/n$ with small $m/n$ a symmetry between the two 
hemisphere takes place. The existence of the clone can be seen clearly 
as a small uniform background at times $t=1/25\,T_{rev}$, $t=1/10\,T_{rev}$. 
There is always a strong interference between the fractional waves 
which does not allow to make a clear counting even for small values of $m/n$.

 \subsection{Elliptic WP} \label{ss5c}
 
For the general elliptic wave packet, as deduced from the previous
discussion there are no clones, but partial revivals with different
topology. Due to the change of symmetry, it is indeed necessary to
make a smooth transition between a system of clones located for
$\eta=1$ in the $Oxy$ plane and a system of rings discussed in
the previous subsection. In the system of coordinates adopted in 
subsection \ref{ss2a} and corresponding to $\eta=0$ these rings
have $Ox$ as symmetry axis. The transition from the clones for $\eta=1$
to the rings for $\eta=0$ is made by developing for $\eta$ smaller
than 1 a system of pairs of crescents perpendicular to the
$Oxy$ plane. This transition is clearly visible for particular
fractional revival times in Fig.~\ref{ringclon}.
For a very small value of $\eta$ the upper part and the
lower part of one crescent meet the corresponding parts of a
symmetric crescent in order to build up such a ring symmetric around
$Ox$. Again this change of topology of this construction forbids the
cloning of all fractional waves. Most of them revive in shapes
different from that of the initial wave packet. We propose to call these
fractional wave packets mutants.
Example of time evolution of wave packet with $\eta=0.5$ is given
in Fig.~\ref{tevoleta}. It is well seen both in Fig.~\ref{ringclon}
and in Fig.~\ref{tevoleta} that clones and mutants can occur
in the same time.
For example for $\eta=\frac{1}{4}$ and $t=\frac{1}{3}\,T_{rev}$
there is a fractional revival at $\theta=\frac{\pi}{2}$ and
$\phi=0$ with the same shape as the initial wave packet and
two crescents which almost close. This situation is similar to that
which exists for $\eta=0$ also for $t=\frac{1}{3}\,T_{rev}$.
For $\eta=\frac{1}{2}$ and  $t=\frac{1}{6}\,T_{rev}$ the two
crescents do not form a ring. Among the three fractional waves
which are present there are two which are identical to each other
but with larger spread in $\theta$. For this value $\eta=\frac{1}{2}$,
there are three different topologies for the five fractional waves.
In other publications on a different system \cite{arvieu}, we
had already found the transition from a gaussian three dimensional
wave packet to a vortex ring. Refs.\ \cite{arvieu} were devoted,
as well as our previous works, for the time evolution of such
coherent waves in the case where the hamiltonian contains a spin orbit
interaction in addition to the harmonic oscillator potential.
If the spin is oriented along the initial wave packet displacement
($Ox$ axis), the cylindrical symmetry is imposed on the system
and preserved during evolution, a vortex rings appear.

The transition between the two shapes of fractional wave packets
can be explained by comparing, as a function of $\eta$,
the spread of the angle $\theta$ to the spread of the angle $\phi$.
If  $\eta=0$ the wave packet has no privileged direction on the
sphere. It spreads therefore equally. In the case when $\eta=1$ the
wave packet is peaked near the plane with $\theta=\frac{\pi}{2}$
and the spread in $\theta$ is strongly reduced. The spread
in $\phi$ is seen in the scenario of cloning in the plane $xOy$.
For intermediate values of $\eta$, there is a competition between
the two effects which manifests itself most strongly in the shape
of the fractional waves.

For $\eta\rangle 1$ (results not presented) we have observed a strong
reduction of the spread in $\theta$. For particular values of
$(m/n)$, some mutants can be peaked with higher amplitude than
their neighbours. This is in direct connection to the increase
in $p_0$ which was necessary in the initial wave packet.

 \subsection{Final remarks} \label{ss5d}
  
  We can measure the aperture of the probability density of the WP by the
solid angle 
$\Omega=4\pi/(4N+1)$ corresponding to the cone defined by
$\tan (\theta'/2)=1/2\sqrt{N}$.
The time of spreading $\tau_\eta$ of this wave packet of width
$\Delta L_\eta$ (\ref{c28}) is of the order of
\begin{equation}\label{c65}	
\tau_\eta = \frac{2\pi}{\omega_0} \, \frac{1}{\Delta L_\eta}  \, .
\end{equation}
The difference in spreading with $\eta$ is clearly seen in the
autocorrelation functions represented in Fig.~\ref{etaucor}.
We can also define
maximum number of fractional wave packets that can be observed
knowing that
$\tan (\theta'/2)=1/2\sqrt{N}$ as
\begin{equation}\label{c66}  
q_{max} = \frac{\pi}{\arctan(1/2\sqrt{N})}   \, .
\end{equation}
This number is confirmed by the observation of the 'carpet' of
Fig.~\ref{carpetM} for $N=20$ for which we have up to 7 clones.
The lifetime ${\tau'}_\eta$ of a system of $q$ fractional wave packets
can be estimated as
\begin{equation}\label{c67}  
\omega_0\,{\tau'}_\eta \, \Delta L_\eta = \frac{2\pi}{q}   \, .
\end{equation}
Also in Fig.~\ref{carpetM} one sees clearly for $q=2,3$ the large
intervals of time during which clones are observed.

\section{Coherent states for a symmetric top} \label{sec6}

It is natural to enlarge our previous study to rotational coherent
states of symmetric tops which contain an additional degree of
freedom. There are several possibilities in the literature for
constructing such a state. Our aim is again to choose the axis of
coordinates in order to eliminate irrelevant parameters and to
make evolution understandable. This is possible only in the case
of symmetric top. We will assume that the moments of inertia in the
intrinsic frame $J_x=J_y$ and introduce the parameter $\delta$
such that:
\begin{equation}\label{delta}	
\delta =  \frac{J_x}{J_z} - 1\, .
\end{equation}
The energy spectrum is then:
\begin{equation}\label{esytop}	
E(I,K) = \frac{\hbar^2}{2J_x}\, [I(I+1)+\delta K^2] \, .
\end{equation}

A convenient rotational wave packet was defined and studied by
Janssen \cite{janssen} on the basis of the work of Perelomov
\cite{perelomov} and Schwinger \cite{schwinger}.
This wave packet denoted by three complex numbers $x, y$ and $z$
is a mixture of $D^I_{MK}$ functions
\begin{eqnarray}\label{xyz}	 
|xyz\rangle & = &\exp(-\frac{1}{2}yy^* (1+xx^*)(1+zz^*)) \\
 & \times & \sum_{IMK}\,
 \sqrt{\frac{(2I)!}{(I+M)!(I-M)!(I+K)!(I-K)!}} \,
 x^{I+M} y^{2I} z^{I+K} \,|IMK\rangle  \, .\nonumber
\end{eqnarray}
The sum over $I$ contains integer as well as half integer values
of $I$ i.e. this state extends that presented by Atkins and Dobson.
On a similar line as in section \ref{sec3} we call $|xyz\rangle_i$
its normalized projection onto the space with integer $I$.
A particularly convenient choice of $x, y$ and $z$ as well as of the
system of coordinates decreases the number of parameters to two and
reduces the summation to $I$ and $K$ only. The simpler wave packet is
then:
\begin{equation}\label{rlamb}  
|r,\lambda\rangle = \exp(-r)\, \sum_{IK} \, (-1)^{I+K}\,(2r)^I \,
\frac{(\sin \frac{\lambda}{2})^{I+K}(\cos \frac{\lambda}{2})^{I-K}}
{\sqrt{(I+K)!(I-K)!}} \, |I-IK\rangle \, .
\end{equation}

If $L_X,L_Y,L_Z$ are the components of $\vec{L}$ in the intrinsic
axes, $L_x,L_y,L_z$ those related to the laboratory axes, one has
according to Janssen:
\begin{equation}\label{jlxyzav} 
\langle r\lambda |L_x|r\lambda \rangle =
\langle r\lambda |L_y|r\lambda \rangle = 0, \hspace{5ex}
\langle r\lambda |L_z|r\lambda \rangle = -r \, ,
\end{equation}
\begin{equation}\label{jlXYZav} 
\langle r\lambda |L_Z|r\lambda \rangle = -r\cos\lambda, \hspace{5ex}
\langle r\lambda |L_X|r\lambda \rangle = -r\sin\lambda, \hspace{5ex}
\langle r\lambda |L_Y|r\lambda \rangle = 0 \, ,
\end{equation}
\begin{equation}\label{jl2av}  
\langle r\lambda |L^2|r\lambda \rangle = r(r+\frac{3}{2}) \, .
\end{equation}

Since the WP given by (\ref{rlamb}) contains only components with 
$M=-I\,$ it is a 
circular WP with $\eta=-1$ which fulfills eq.\ (\ref{c2}) with the 
components of $\vec{L}$ taken in the laboratory system.
From Janssen's work the components in the 
rigid body system verify the equation 
\begin{equation}\label{body1}  
           \Delta L_X^2 = \Delta L_Y^2  \; ,
\end{equation}
while their product takes the value
\begin{equation}\label{body2}  
         \Delta L_X^2 \, \Delta L_Y^2 = 
	   \frac{\langle L_Z \rangle ^2}{4\cos ^2\lambda} \; .
\end{equation}
Therefore eq.\ (\ref{c2}) is also verified for the components in the 
rigid body system if $OZ$ is directed along $\vec{L}$.
On the other hand 
the expansion (\ref{rlamb}) of the rotational coherent states in terms
of $|I-IK\rangle$ has the same coefficients as the expansion of
Atkins--Dobson in terms of angular momentum eigenstates $|IK\rangle$.
The two expansions are connected by defining $\alpha_+$ and
$\alpha_-$ as:
\begin{equation}\label{alpha}  
\alpha_+=\sqrt{2r}\,\sin \frac{\lambda}{2}, \hspace{10ex}
\alpha_-=\sqrt{2r}\,\cos \frac{\lambda}{2} \, .
\end{equation}

The projection $|r,\lambda\rangle_i$ deduced by restricting
(\ref{rlamb}) to integer values of $I$ and $K$ verify eqs.
(\ref{jlxyzav}) to  (\ref{jl2av}) only approximately. However if
$r$ is large enough, these equations are obtained in a very good
approximation. The time evolution of the wave packet
$|r,\lambda\rangle_i$ will be studied in the next section for a rigid
body symmetric rotor.

\section{Time evolution of Janssen's coherent state} \label{sec7}

The energy spectrum of the axially symmetric rigid rotor is written as:
\begin{equation}\label{esumr}  
E_{IK} = \hbar\omega_0 \, [I(I+1) + \delta K^2] \,.
\end{equation}
We will study the time evolution of a wave packet deduced from
(\ref{rlamb}) with $\omega_0=\hbar/(2J_x)$ and with average values
\begin{eqnarray}\label{avLZi} 
\langle r\lambda |L_Z|r\lambda \rangle &=&\bar{K}\simeq -r\cos\lambda\\
\label{avLzii} 
\langle r\lambda |L_z|r\lambda \rangle &=& \bar{I} \simeq -r \, .
\end{eqnarray}
The energy $E_{IK}$ is written by taking $\bar{K}$ and $\bar{I}$ as
reference, and defining $k_1$ and $k_2$ as:
\begin{equation}\label{k1k2}   
k_1= I - \bar{I} \hspace{10ex} k_2= K - \bar{K} \, .
\end{equation}
\begin{equation}\label{ek1k2}	
E_{k_1 k_2} = \hbar\omega_0\,[\bar{I}(\bar{I}+1) + \delta \bar{K}^2]
+ \hbar\omega_0\,(2\bar{I}+1)k_1 + \hbar\omega_0\,\delta(2\bar{K})k_2
+ \hbar\omega_0\,k_1^2 + \hbar\omega_0\,\delta\,k_2^2 \, .
\end{equation}
We now follow the lines drawn by Bluhm, Kostelecky and Tudose
\cite{bluhm1}
who have considered the time evolution of a system which depends
quadratically on two quantum numbers, in our case $I$ and $K$.
There are 4 time constants; the first pair defined as
\begin{equation}\label{tclrI}	
T_{cl}^I = \frac{2\pi}{\omega_0(2\bar{I}+1)} \hspace{10ex}
T_{rev}^I = \frac{2\pi}{\omega_0}=(2\bar{I}+1)\,T_{cl}^I
\end{equation}
is related to the motion around $Oz$ axis (laboratory axis),
i.e. it is connected to the Euler angle $\alpha$, the second pair
plays a parallel role, it concerns the motion around the symmetric
$OZ$ axis and the Euler angle $\gamma$
\begin{equation}\label{tclrK}	
T_{cl}^K = \frac{2\pi}{\omega_0\delta 2\bar{K}} = \frac{1}{\delta}
\frac{2\bar{I}+1}{2\bar{K}}\, T_{cl}^I
\hspace{10ex}
T_{rev}^K = \frac{2\pi}{\delta \omega_0}=\frac{1}{\delta}\,T_{rev}^I
\, .
\end{equation}

The system of coordinates and the parametrization in (\ref{rlamb})
enables to profit by the separation of variables in the state
$|I-IK\rangle$ since:
\begin{equation}\label{abgIIK}	 
\langle\alpha\beta\gamma|I-IK\rangle = e^{i\alpha I}\,
d_{-IK}^{\,I}(\beta)\, e^{-i\gamma K} =
D_{-IK}^{\,I}(\alpha,\beta,\gamma) \, .
\end{equation}

The wave packet (\ref{rlamb}) at time $t$ with conditions
(\ref{avLZi}) and (\ref{avLzii}) and integer $I$ and $K$ will be
denoted as $|\bar{I}\,\bar{K}\rangle$ and written as:
\begin{eqnarray}\label{abgIKt}	 
\langle\alpha\beta\gamma|\bar{I}\,\bar{K}\rangle_t
&=& \sum_{IK}\, C_{IK}(r,\lambda)\, d_{-IK}^{\,I}(\beta) 
e^{i[\alpha I -2\pi I(I+1)\,t/T_{rev}^I]}\,
e^{-i(\gamma K +2\pi K^2\, t/T_{rev}^K)}  \\
\label{abgIKt1} 		  
&=& e^{-i[\bar{I}(\bar{I}+1)+\delta \bar{K}^2]\, t/T_{rev}^I} \,
\sum_{k_1 k_2} \, C_{k_1 k_2}(r,\lambda) \, d^{\,I}_{k_1 k_2}(\beta) \\
& \times &  e^{i[\alpha k_1-2\pi(k_1/T_{cl}^I+k_1^2/T_{rev}^I)t]}\,
e^{-i[\gamma k_2+2\pi(k_2/T_{cl}^K+k_2^2/T_{rev}^K)t]}
\nonumber \, .
\end{eqnarray}
The summation on $I$ and $K$ has been changed to a sum over $k_1$ and
$k_2$ and the coefficient $C_{IK}\,d_{-IK}^{\,I}$ has been given the new
indexes. The discussion of the time evolution of (\ref{abgIKt}-\ref{abgIKt1})
follows in a straightforward manner that given by Bluhm,
Kostelecky and Tudose \cite{bluhm1}.
The crucial parameter is $\delta$. If $T_{rev}^K$ and $T_{rev}^I$
are not commensurate there is no cloning, however for
$t=(m/n)T_{rev}^I$ there are partial revivals in the variable
$\alpha$: i.e. the wave packet is a superposition of $q$ fractional
wave packets
peaked regularly along the $Oz$ axis. ($q=(n/2)$ if $n$ is even,
$n$ in other cases). For $t=(m'/n')T_{rev}^K$ the same scenario
of partial revival occurs but this time there are $q'$ fractional wave
packets
peaked around the $OZ$ axis. [$q'=(n'/2)$ if $n'$ is even,
$n'$ otherwise].

The interesting situation of commensurability of $T_{rev}^I$ and
$T_{rev}^K$ allows, on the contrary, the construction of $q^2$
clones for all the time such that
\begin{equation}\label{tclone}	 
t = \frac{m}{n}\,T_{rev}^{I,K}
\end{equation}
The revival time $T_{rev}^{I,K}$ is the least common multiple of
$T_{rev}^I$ and $T_{rev}^K$
\begin{equation}\label{tcomrev}  
p\,T_{rev}^I= r\,T_{rev}^K = T_{rev}^{I,K} \, .
\end{equation}
The time evolution of the rotational wave packets is presented in
Fig.~\ref{tevoljir} and Fig.~\ref{tevoljr}.

The particular choice $\bar{K}=0$ leads to $T_{cl}^K = \infty$
but $T_{rev}^K$ is finite as well as $T_{cl}^I$ and $T_{rev}^I$.
For the smaller value of $t$, the behaviour of the wave packet around
$Oz$ and $OZ$ is different as seen from Fig.~\ref{tevoljr}.
There is indeed a classical rotation and spreading around $Oz$ while
no rotation occurs around $OZ$, only the spreading is observed around
this axis.

Fig.~\ref{tevoljir}  illustrates the case of an irrational
value of $\delta$
where there are partial revivals for times $t=(m/n)\,T_{rev}^I$
around $Oz$ and for times $t=(m'/n')\,T_{rev}^K$ where the
revivals are around $OZ$. Note that the concentration of the
wave packet for some definite values of $\alpha$ (or $\gamma$)
has no influence on the other variable $\gamma$ (or $\alpha$)
respectively.

Fig.~\ref{tevoljr}  illustrates the case of a rational value of
$\delta$,
for which there are $q^2$ clones. The proof that there are $q^2$ clones
if condition (\ref{tcomrev}) holds is rather simple, one applies twice
the method of Averbukh and Perelman to linearize the exponential
containing $k_1^2$ and $k_2^2$ in eq. (\ref{abgIKt}-\ref{abgIKt1}).

Finally we want to stress that the variable $\beta$ does not play
a role in the time evolution. Obviously this is due to our
particular choice of axis in eq. (\ref{rlamb}). The role played
by $\beta$ for the axial rotor is similar to the role played by
$\theta$ in the diatomic molecule. 

\section{Conclusions} \label{sec8} 

We have discussed the time evolution of angular momentum
coherent states that are built from intelligent spin states,
but as pointed in the Appendix, can be constructed by a
simple, purely geometric, generating procedure.
The basic ingredients are the three dimensional gaussian wave packets
of the harmonic oscillator. We have used two of the parameters
of the gaussian to derive an ensemble of coherents states
which depend finally on two parameters $N$ and $\eta$.
The angular distribution of the probability density depends only
on $N$, while the momentum distribution depends both on $N$ and $\eta$.
The uncertainty relation (\ref{c2}) is valid for all
$N$ and $\eta$. Such a variety of wave packets do not arise from
the previous works on angular momentum coherent states with bosons.
It is for example simple to show \cite{fonda}
that there is no place in the
Atkins and Dobson or Mihailov's work for the wave packets (\ref{c35}-\ref{c36})
which correspond to $\eta=0$. Also
the coefficient $\alpha_{jm}$ defined by
(\ref{c40}) in terms of $\alpha_{+}$ and $\alpha_{-}$ cannot
be put generally in the form such that (\ref{c39})
which ensures that  $b_{IM}$
is zero if $I$ and $M$ have an opposite parity.

In the framework of the scenario of ref.~\cite{averbukh}
we have shown that for $\eta=1$ the wave packets spread at
$t=\frac{m}{n}\,T_{rev}$ into clones. However, the difference
between the symmetries for $\eta=0$ and $\eta=1$ necessitates
a change in the topology of the fractional wave packets.
This change is such that these wave packets can be named mutants
after a certain amount of deformation of their shape.
Aronstein and Stroud \cite{aronstein} have shown "that, the
infinite square well is an ideal system for fractional revivals
since all wave functions exhibit fractional revivals of all orders".
The same statement applies to the case of the diatomic molecule
but is limited however to circular wave packets, i.e. to those
which contain only $M=I$ partial waves in a convenient system of axis.
The generation mechanism used in this paper can be extended to shapes
different from gaussian distribution. Clones will always result
in this particular condition. For the wave packets which contain
populations of sublevels $M\neq I$, mutants will appear. All these
results hold independently on the realization of the uncertainty
condition (\ref{c2}). For systems where energy is quadratic in
one quantum number we have shown that the time evolution can
exhibit a rich structure of cloning and mutation of fractional
revivals.
It is a challenging question to find a proper way to excite
such wave packets in a rotational band of a molecule or
of a deformed nucleus. 
Let us briefly describe the efforts made under this perspective. 

The wave packets with $\eta=0$ or wave packets with $M=0$
have been already considered several times. In a reference
work on Coulomb excitation of nuclei with heavy ions
R.A.~Broglia and  A.~Winther \cite{broglia} have shown that
such wave packets are generated in  backward scattering.
Their population of the excited states of a rotational
band with $K=0$ do not coincide with the very
simplified expression (\ref{c35}-\ref{c36}).
Indeed interference effects during the excitation process
modulates this population to a large extent (Fig.~8, p.82 of
ref. \cite{broglia}). However, we have studied the time
evolution of such a wave packet and this will
be the subject of a future publication \cite{arvieu1}.

The authors of ref.\ \cite{fonda} have written an important analysis
on coherent rotational states, completing the work of \cite{broglia}.
They have already used the wave $|k,\frac{1}{2}\rangle_i$ of
section \ref{sec3} derived from \cite{atkins},
and they have also considered an extension toward other symmetries
like Sp(2,R), Sp(4,R). Time evolution was studied but only
for the average values of operators like quadrupole moments.
The physical systems considered were the nucleus $^{238}$U and
the molecule CS$_2$. The time evolution of expectation values
of more general operators for wave packets due to quadratic
dependence was studied in \cite{braun}.
The operators studied in ref.\ \cite{fonda} have very strict
selection rules and the time evolution reflected there is much
poorer than what is exhibited by the full wave packet and by the
average values of the more complex observables	\cite{braun}
related to the position.
It is, however, clear that all the wave packets considered in
ref.\ \cite{fonda} must evolve according to the scenario of ref.\
\cite{averbukh} with a very rich sequence of changes of shapes.

   Our work comes very near from the spirit of two recent publications 
devoted to molecular physics. In ref.\ \cite{persico} the population of a set of 
rotational states of a molecule by an intense laser has been 
calculated. However, the number of states excited in this way is rather 
small, $l<4$, the symmetry implies only states with $M=0$ and therefore the WP 
produced is not concentrated on the sphere as sharply as ours with $N=20$ 
or 50.
   In ref.\ \cite{ortigoso} Ortigoso has studied how to tailor microwaves pulses in
order to create rotational coherent states for an asymmetric--top
molecule, which are built from Radcliffe's intelligent spin states
discussed in section \ref{sec2}. It is gratifying that such a mechanism which 
involves optimal control theory is possible. However our demand is more 
ambitious since one needs to combine intelligent spin states of different 
angular momentum in order to achieve a proper angular concentration. In 
addition one has to find a mechanism that allows to change continuously 
our variable $\eta$, i.e.\ the relative uncertaities in $L_x$ and $L_y$,
in order to 
explore the whole variety of our states as well as of the intelligent 
spin states. It is indeed interesting that Radcliffe's states have been 
used quite thorougly for example in ref.\ \cite{huber,martens}  
but our paper points 
toward the fact that a richer structure exists nearby in accordance to 
older works \cite{aragone1,aragone2,rashid}. 
As said in the text we have not used the larger class 
of intelligent spin states in the case of the diatomic molecule and we 
are also conscious that our work on the symmetric top leaves open 
possibilities of coherent states that have not yet been explored.

It is worth to mention very recent paper by Chen and Yeazell
\cite{chen} on an analytical wave packet design scheme that is able
to create desired Rydberg wave packets and control their
dynamics. 

   We believe that we have enriched anyway the exemples given in the
reference \cite{bluhm1}  
by pointing out how different the fractional revivals may be.

\appendix
\section{Connection between the exponential coherent WP and 
coherent states of the harmonic oscillator}
  
      The exponential WP defined by eq.\ (\ref{c18}) can be
manufactured from gaussian 
wave packets in three dimensions which are coherent states of the 
harmonic oscillator. Such a general WP with an average position 
$\vec{r}_0$ and momentum $\vec{p}_0$ are written as
\begin{equation} \label{a1} 
\Psi_G(r,\theta,\phi)= \frac{1}{(2\pi)^{\frac{3}{4}}
\sigma^{\frac{3}{2}}} \,
\exp{\left [-\,\frac{(\vec{r}-\vec{r_0})^2}{2\sigma^2}
+\,i \,\frac{\vec{p_0}\cdot\vec{r}}{\hbar} \right ]} \, .
\end{equation}
 One gets rid of three unecessary parameters if one chooses the axis in 
such a way that
\begin{equation} \label{a3} 
 {p_0}_z = 0 ,\hspace{5ex} \vec{r}_0 = \hat{x}\,r_0 \; .
\end{equation}
With such choices the probability density is
\begin{equation} \label{a4} 
 |\Psi_G|^2 = \left [ \frac{1}{(2\pi )^{3/2}\,\sigma^3}  
 \exp \left (-\frac{r^2+r_0^2}{\sigma^2}\right ) \right ] \,
 \exp \left (\frac{2r\,r_0}{\sigma^2}\,\sin \theta\,\cos \phi \right ) \; .
\end{equation}
 Apart from a normalization factor it coincides with eq.\ (\ref{c19}) 
 if $r$ and $r_0$ are 
both chosen as 
\begin{equation} \label{a5} 
          r = r_0=\sigma \,\sqrt{N}  \; .
\end{equation}
  The solid angle $\Omega=4\pi/(4N+1)$ is thus the angle with 
which the width $\sigma$ of the density of the gaussian WP is observed 
from the center of the sphere of radius given by (\ref{a5}). 
 This choice still leaves two parameters ${p_0}_x$ and ${p_0}_y$ free.
 The choice which leads to (\ref{c18}) is 
\begin{equation} \label{a6} 
          {p_0}_x=0\, ,\hspace{5ex}   r_0\,{p_0}_y = \eta\,N \,\hbar  \; .    
\end{equation}
    If one takes the width of the harmonic oscillator
$\sigma=\sqrt{\hbar/m\,\omega}$ the value of ${p_0}_y$ given in (\ref{a6}) 
leads for $\eta=1$ to 
\begin{equation} \label{a7} 
       {p_0}_y = m\, r_0 \, \omega    \; .  
\end{equation}
   The coherent state (\ref{a1})  associated to this value   
evolves around a circular trajectory in the field of the harmonic oscillator.
 In the same manner the value $\eta=0$ is associated to 
a linear trajectory and the other values of $\eta$ correspond 
to elliptic trajectories. In the second case the initial 
point is either the apogee or the perigee of the ellipsis. In this manner 
more general WP can also be constructed which start from arbitrary points 
according to a nonzero value given to ${p_x}_0$. Let us define a parameter 
$\epsilon$ by 
\begin{equation} \label{a8} 
    r_0 \,{p_0}_x  =  \epsilon  N\hbar    \; .  
\end{equation}
 The exponential WP (\ref{c18}) becomes in these new conditions
\begin{equation} \label{a9} 
  \Psi_{\eta,\epsilon} (\theta,\phi) = \sqrt{\frac{N}{2\pi\sinh 2N}} \,
   \mbox{e}^{N\sin\theta[(1+i\epsilon)\cos\phi +i\eta\sin\phi]}    
\end{equation}
and verify the condition 
\begin{equation} \label{a10} 
[(1+\mbox{i}\epsilon) L_x +\mbox{i}\eta L_y]\,\Psi_{\eta,\epsilon} =0 \;. 
\end{equation}
 The probability density is again given by (\ref{c19}). However these more 
general states do not fulfill condition (\ref{c2}). 
Indeed for those states which obey this equation there exists a value 
of $\eta$ given by (\ref{c7}) such that 
eq.\ (\ref{c6}) is obeyed and eq.\ (\ref{a10}) is not.
It was then consistant to consider only those WP with $\epsilon=0$.
This result is in accordance with an older result by Rashid \cite{rashid}
and mark a difference between so called {\em quasi--intelligent spin states}
which solve (\ref{a9}) but not satisfy (\ref{c2}) and the 
{\em intelligent spin states} for which $\epsilon =0$.


\newpage

\begin{figure}[bh]
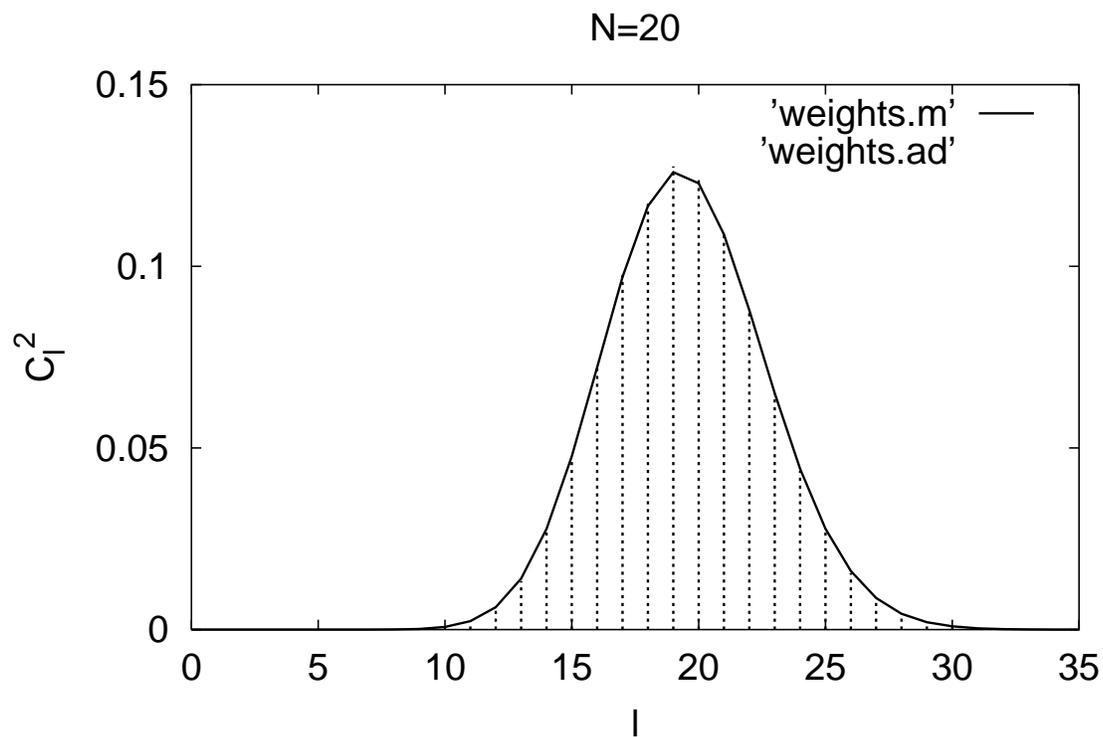
 
\caption[wagi]{Probabilities to find the partial waves $Y_I^I$
in the coherent states of Mostowski (solid line) and Atkins--Dobson
(dashed impulses) for parameters $N=20=1/2(k^2+1)$ ensuring the same
angular velocity for both wave packets.}
\label{wagima}
\end{figure}

\begin{figure}[bth]
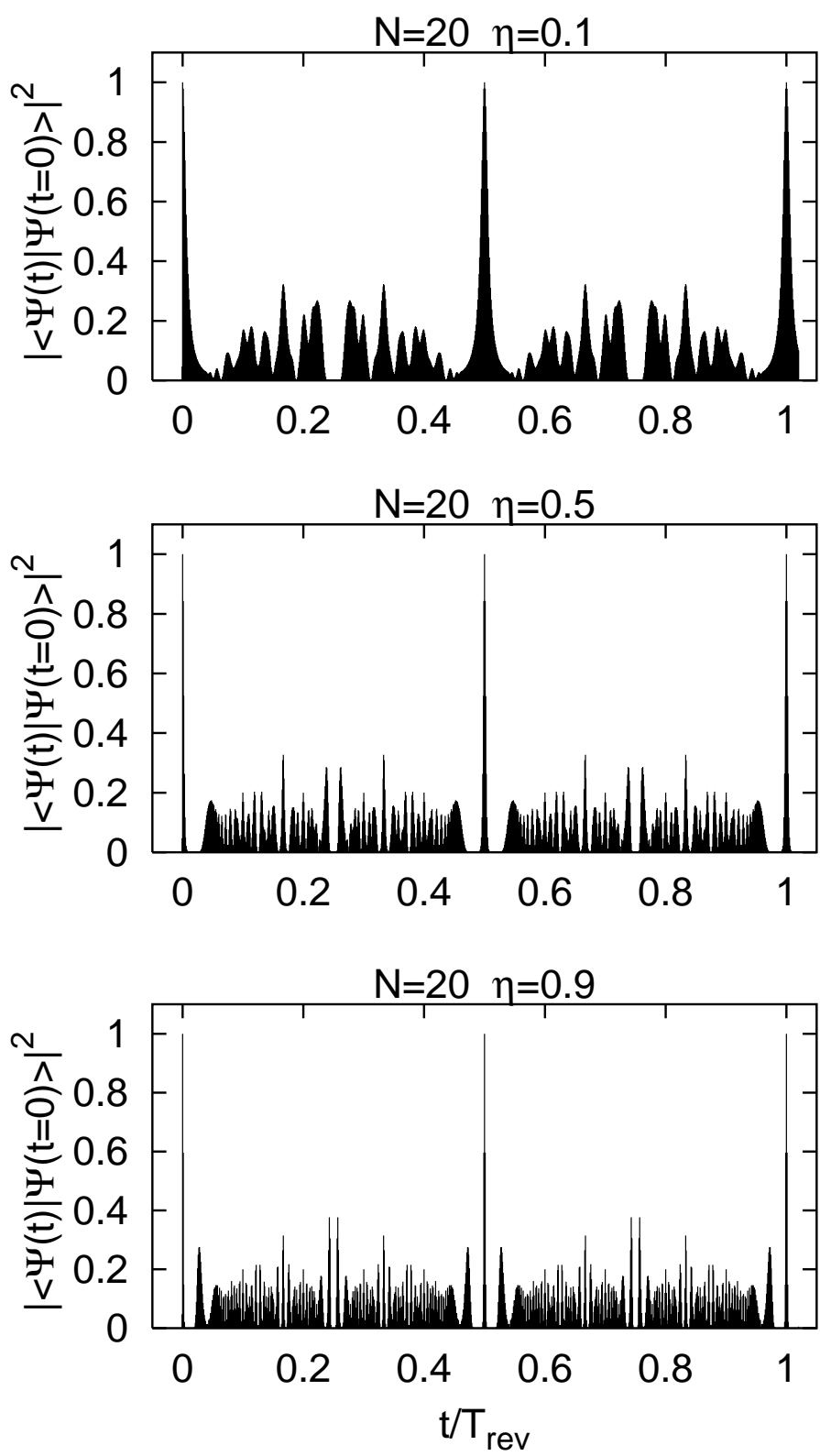

\caption[aucoreta]{The autocorrelation function for $N=20$
and different values of parameter $\eta$ corresponding
to a smooth transition between two different symmetries.}
\label{etaucor}
\end{figure}

\begin{figure}[h]
\caption[Mevol]{Time evolution of Mostowski's wave packet
 with $N=20$.  The left column
represents changes of the probability density during short term
evolution, the right one at fractional revival times.
The probability density at times equal 1/5, 1/3 and 1/2*$T_{rev}$
are identical with those presented at 1/10, 1/6 and n*$T_{rev}$,
respectively. (The vertical scale is not the same in all figures.)}
\label{tevolM}
\end{figure}

\begin{figure}[h]
\caption[Mcarpet]{Time evolution of Mostowski's wave packet with
$N=20$. The $|\Psi_M(\theta,\phi,t)|^2$ for fixed $\theta=\pi /2$
is presented in the contour plot. The larger values of
 $|\Psi_M(\theta,\phi,t)|^2$ result in heavy lines due to almost
 overlaping cuts for fractional revivals. One can clearly see
 the fractional revivals of orders 1/7, 1/12, 1/5, 1/8, 1/6, 1/4
 and so on (corresponding times are 1/7, 1/6, 1/5, 1/4, 1/3 and
 1/2 of $T_{rev}$). }
\label{carpetM}
\end{figure}

\begin{figure}[bh]
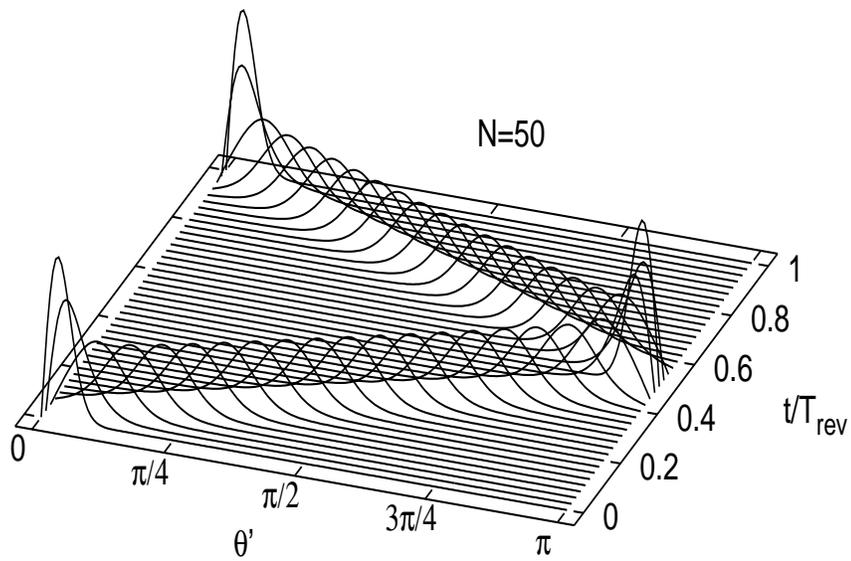

\caption[Lclass]{Time evolution of the $\Psi_{cl}$ (\ref{c64})
for $N=50$. The probability density $2\pi\sin\theta'|\Psi_{cl}|^2$
is presented as a function of $\theta'$ and $t$ within
one revival period.}
\label{classL}
\end{figure}

\begin{figure}[h]
\caption[Lcarpet]{Time evolution of the wave packet (\ref{c30}-\ref{c31})
for $N=50$. 
The probability density $2\pi\sin\theta'|\Psi|^2$
is presented in the contour plot. The larger values of
 $2\pi\sin\theta'|\Psi|^2$ result in heavy lines due to almost
 overlaping cuts for fractional revivals.}
\label{carpetL}
\end{figure}

\begin{figure}[bh] 
\caption[etanevol]{Time evolution of the WP (\ref{psietan}) 
with infinite $\eta$ but $\eta N=20$.
At $t=0$ the probability density is $1/4\pi$. 
See explanations in the text.}
\label{evetan}
\end{figure}

\begin{figure}[bth]
\caption[clonring]{Transition of fractional wave packets
from exact clones ($\eta=1$) through developing crescents
($\eta=1/2,\eta=1/4$) to ring topology ($\eta=0$) is demonstrated
for two fractional revival times $t=1/3*T_{rev}$ (left)
and $t=1/4*T_{rev}$ (right). The fractional waves called mutants
are clearly seen in the lower rows of the figure.}
\label{ringclon}
\end{figure}

\begin{figure}[bth]
\caption[etaevol]{Time evolution of the coherent state
deduced from elliptic motion ($\eta=0.5$).
The time sequence presented here is the same as in Fig.\ \ref{tevolM}.
Comparing to this figure the crescents are clearly visible,
they are the most pronounced for $t=\frac{1}{4}\,T_{rev}$.}
\label{tevoleta}
\end{figure}

\begin{figure}[bh]
\caption[Jevol]{Time evolution of Janssen's coherent state
for axially symmetric top (\ref{abgIKt}-\ref{abgIKt1}) 
for irrational $\delta=1/\sqrt{3}$ implying
$T_{rev}^I=1/\sqrt{3}\,T_{rev}^K\simeq 0.577\,T_{rev}^K$.
Shown is the probability density
$|\langle\alpha\beta\gamma|\bar{I}\bar{K}\rangle|^2$ for
$\beta=\pi/2$, $\bar{I}=4$ and $\bar{K}=0$.
Notice that the vertical scale is not the same in all figures.}
\label{tevoljir}
\end{figure}

\begin{figure}[bh]
\caption[Jevol]{The same as in Fig.~\ref{tevoljir}
but for rational $\delta=1/2$
implying $T_{rev}^{I,K}=T_{rev}^{K}=2\,T_{rev}^{I}$.
Clones for times 1/6, 1/3 and 1/2$\,T_{rev}^{I,K}$ are clearly
visible as well as full revivals.
Notice that the vertical scale is not the same in all figures.}
\label{tevoljr}
\end{figure}
\vfill

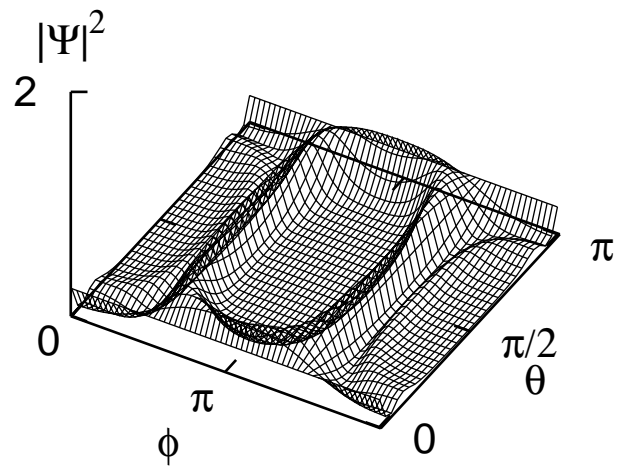
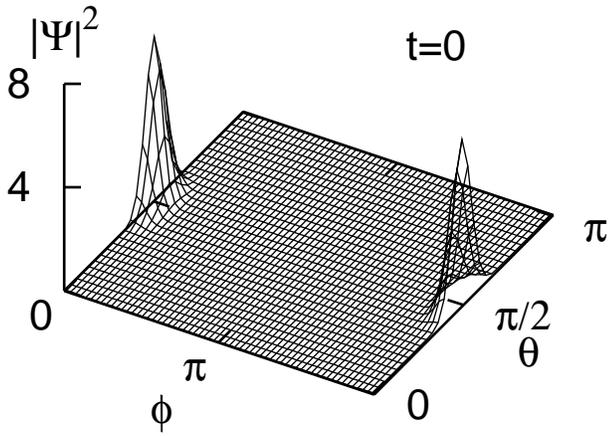
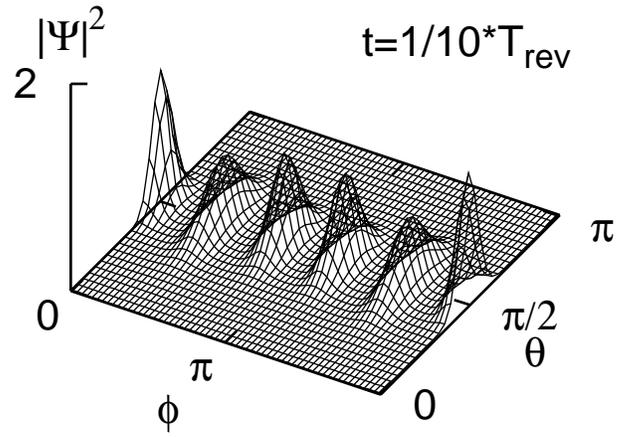
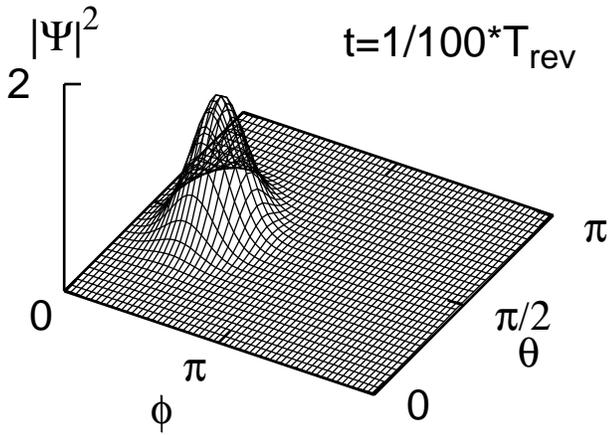
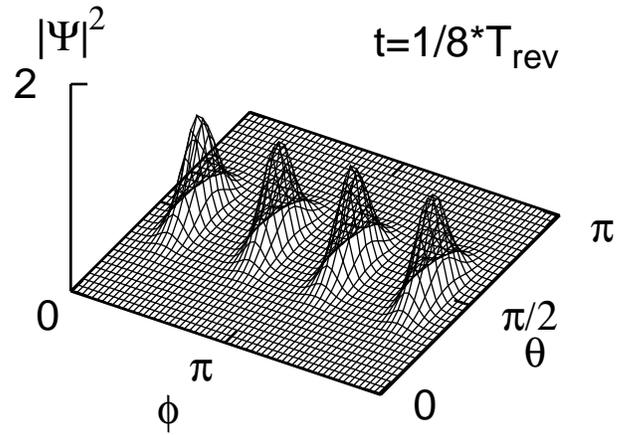
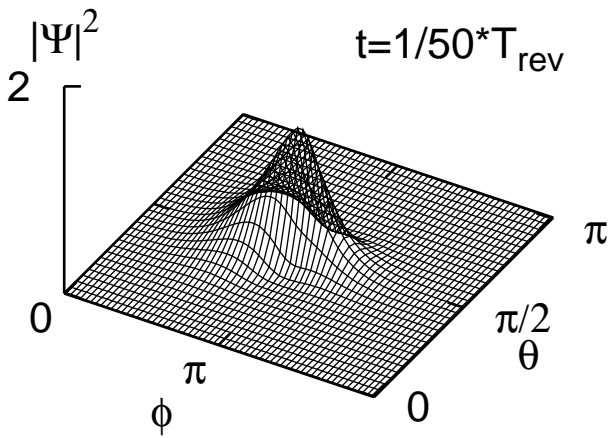
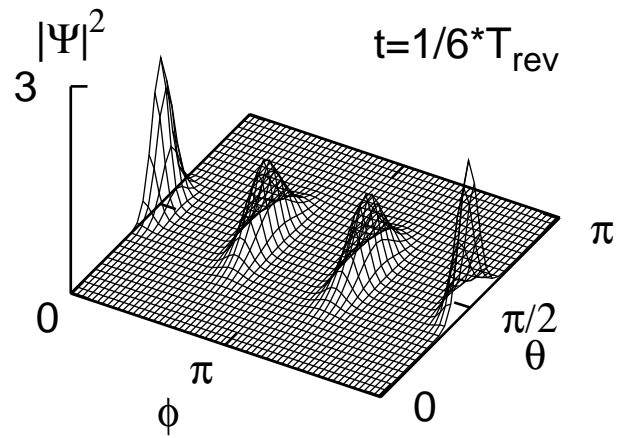
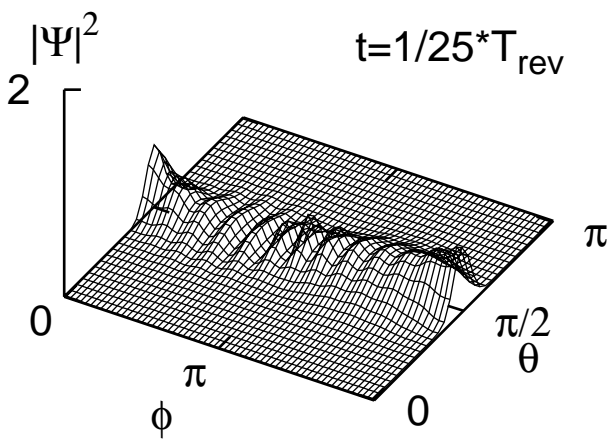
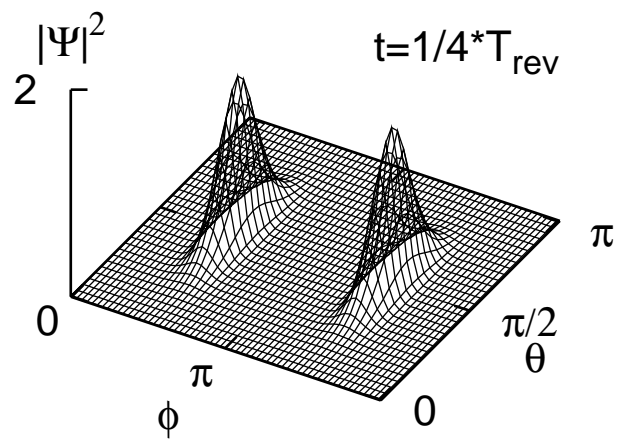
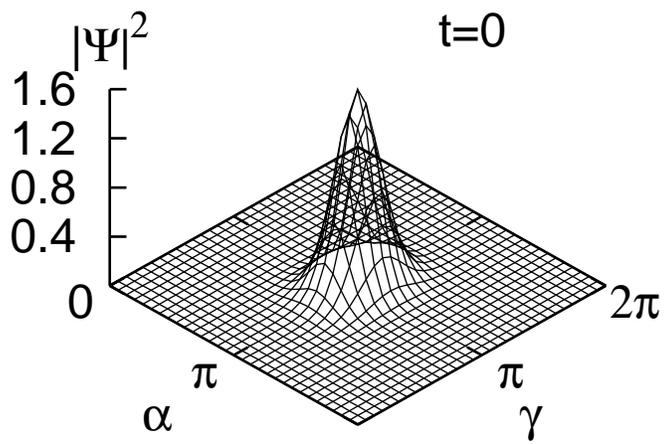
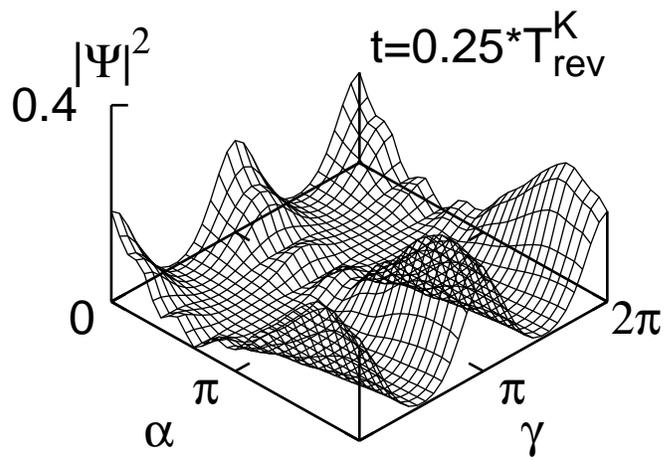
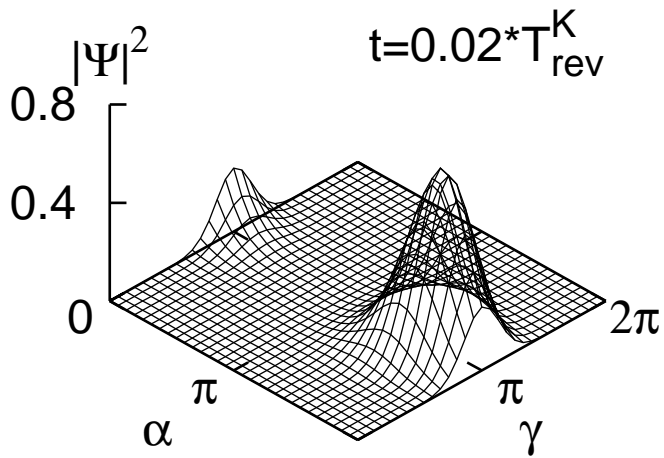
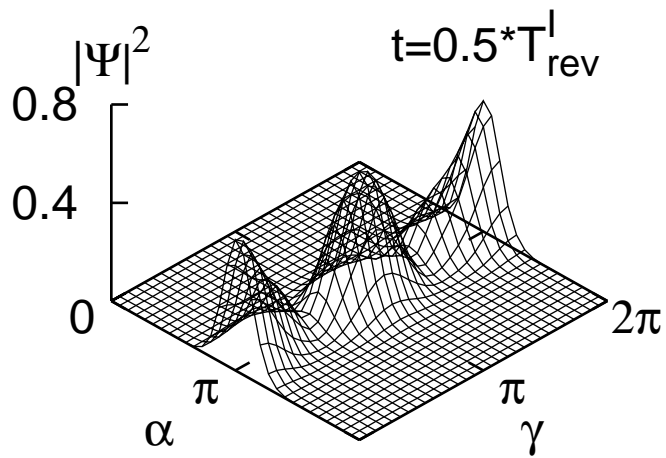
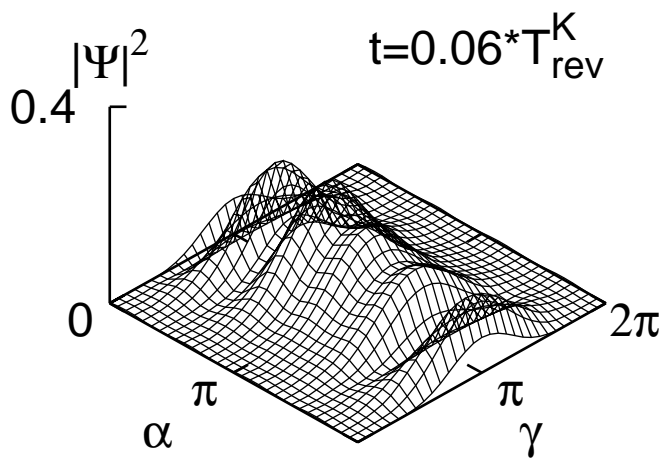
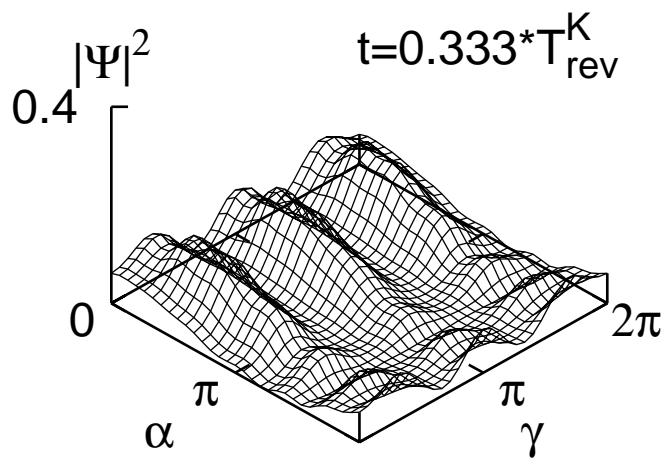
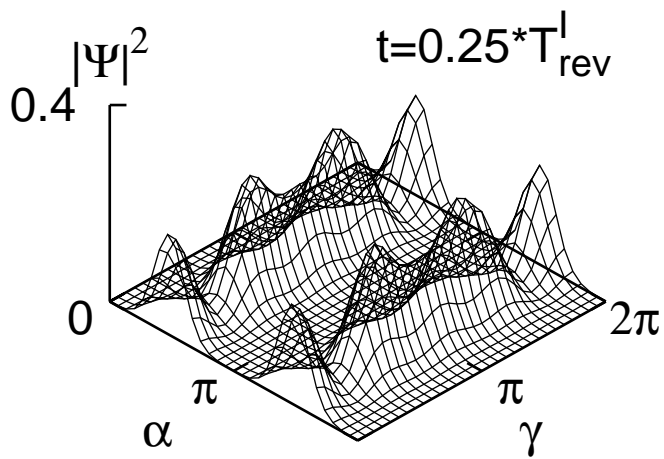
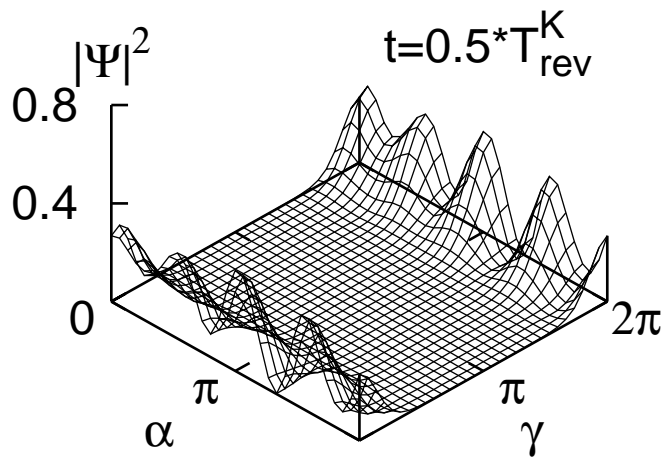
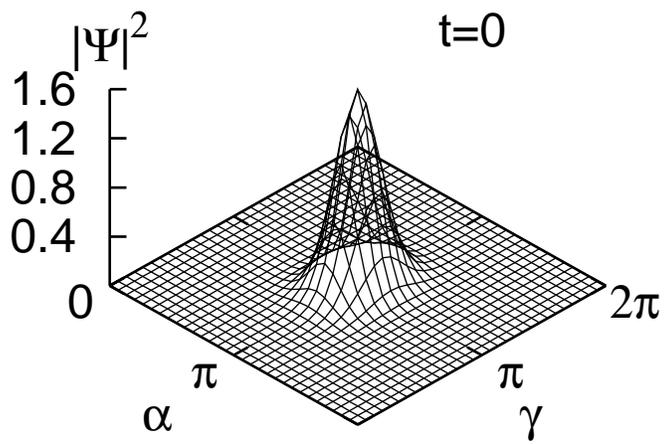
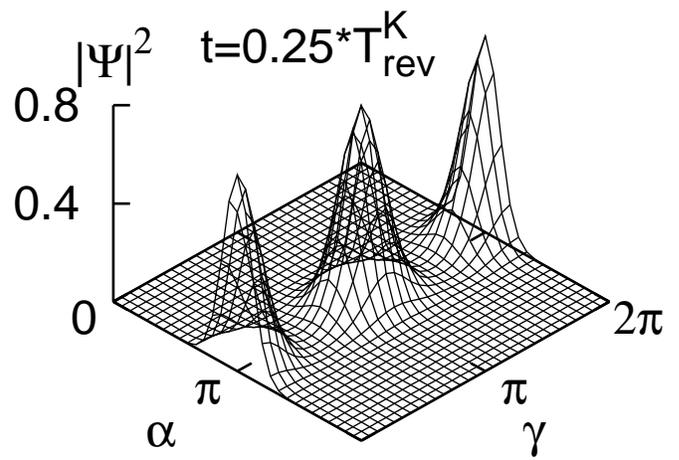
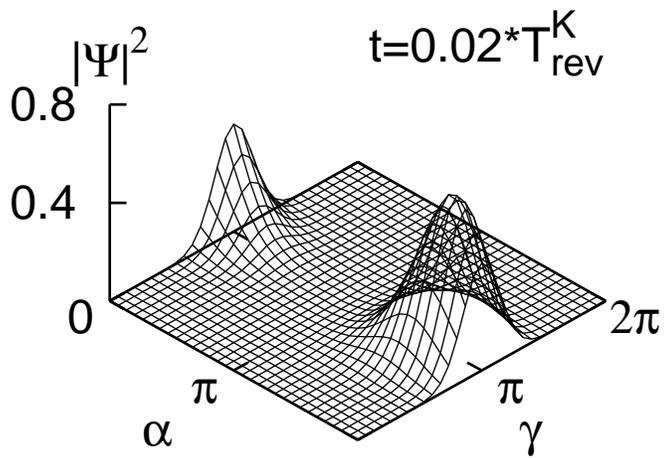
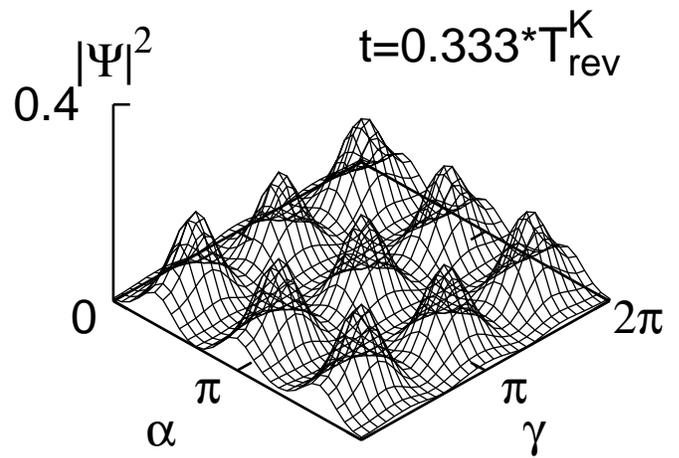
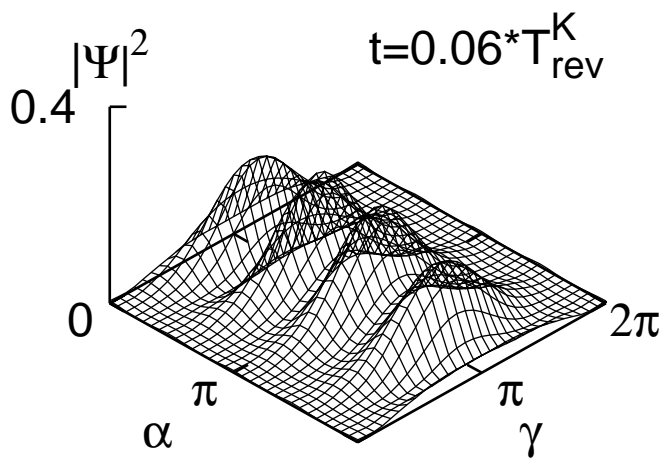
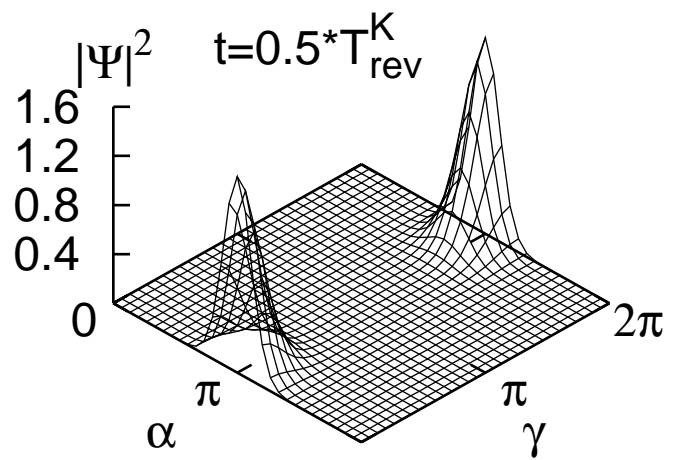
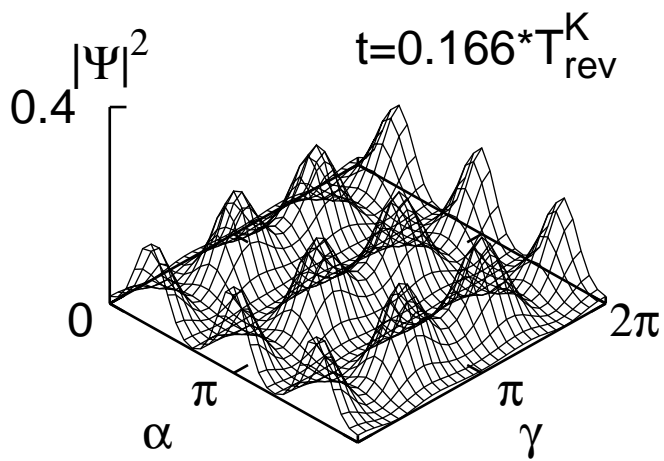
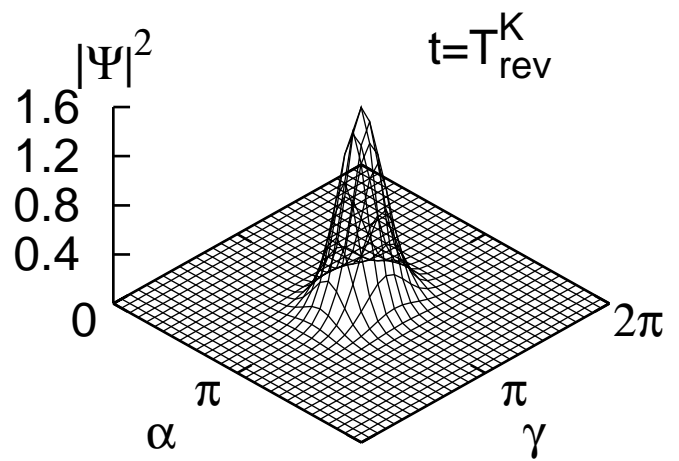
\end{document}